\documentclass[acmsmall,screen]{acmart}
\usepackage{amsmath,mathtools}
\usepackage[ruled,vlined]{algorithm2e}
\usepackage{multirow}
\usepackage{graphicx}
\usepackage{subfigure}
\usepackage{caption}
\usepackage{booktabs}
\usepackage{xcolor,colortbl}
\usepackage{etoolbox}
\usepackage{listings}
\definecolor{purple}{HTML}{eae5f5}

\lstdefinelanguage{Markdown}{
    basicstyle=\linespread{0.8}\ttfamily\footnotesize, 
}
\AtBeginDocument{%
  \providecommand\BibTeX{{%
    \normalfont B\kern-0.5em{\scshape i\kern-0.25em b}\kern-0.8em\TeX}}}

\setcopyright{acmlicensed}
\acmJournal{PACMHCI}
\acmYear{2024} \acmVolume{8} \acmNumber{CSCW1} \acmArticle{163} \acmMonth{4}\acmDOI{10.1145/3641002}

\begin{document}
\title[Embedding Democratic Values into Social Media AIs]{Embedding Democratic Values into Social Media AIs via Societal Objective Functions}

\author{Chenyan Jia}
\email{chenyanj@stanford.edu, c.jia@northeastern.edu}
\authornote{Both authors contributed equally to this research.}
\orcid{0000-0002-8407-9224}
\affiliation{%
  \institution{Stanford University, USA; Northeastern University}
  \streetaddress{450 Jane Stanford Way}
  \city{Stanford}
  \state{California}
  \country{USA}
  \postcode{94305}
}
\author{Michelle S. Lam}
\authornotemark[1]
\email{mlam4@cs.stanford.edu}
\orcid{0000-0002-3448-5961}
\affiliation{%
  \institution{Stanford University}
  \city{Stanford}
  \state{CA}
  \country{USA}
}
\author{Minh Chau Mai}
\email{mcmai@stanford.edu}
\orcid{0000-0002-2273-1256}
\affiliation{%
  \institution{Stanford University}
  \city{Stanford}
  \state{CA}
  \country{USA}
}
\author{Jeffrey T. Hancock}
\email{hancockj@stanford.edu}
\orcid{0000-0001-5367-2677}
\affiliation{%
  \institution{Stanford University}
  \city{Stanford}
  \state{CA}
  \country{USA}
}
\author{Michael S. Bernstein}
\email{msb@cs.stanford.edu}
\orcid{0000-0001-8020-9434}
\affiliation{%
  \institution{Stanford University}
  \city{Stanford}
  \state{CA}
  \country{USA}
}

\renewcommand{\shortauthors}{Chenyan Jia et al.}

\begin{abstract}
Mounting evidence indicates that the artificial intelligence (AI) systems that rank our social media feeds bear nontrivial responsibility for amplifying partisan animosity: negative thoughts, feelings, and behaviors toward political out-groups. 
Can we design these AIs to consider democratic values such as mitigating partisan animosity as part of their objective functions?
We introduce a method for translating established, vetted social scientific constructs into AI objective functions, which we term \textit{societal objective functions}, and demonstrate the method with application to the political science construct of anti-democratic attitudes.
Traditionally, we have lacked observable outcomes to use to train such models --- however, the social sciences have developed survey instruments and qualitative codebooks for these constructs, and their precision facilitates translation into detailed prompts for large language models. We apply this method to create a \textit{democratic attitude model} that estimates the extent to which a social media post promotes anti-democratic attitudes, and test this democratic attitude model across three studies.
In Study 1, we first test the attitudinal and behavioral effectiveness of the intervention among US partisans ($N = 1,380$) by manually annotating ($\alpha = .895$) social media posts with anti-democratic attitude scores and testing several feed ranking conditions based on these scores. Removal ($d = .20$) and downranking feeds ($d = .25$) reduced participants' partisan animosity without compromising their experience and engagement. In Study 2, we scale up the manual labels by creating the democratic attitude model, finding strong agreement with manual labels ($\rho = .75$). Finally, in Study 3, we replicate Study 1 using the democratic attitude model instead of manual labels to test its attitudinal and behavioral impact ($N = 558$), and again find that the feed downranking using the societal objective function reduced partisan animosity ($d = .25$). This method presents a novel strategy to draw on social science theory and methods to mitigate societal harms in social media AIs.

\end{abstract}

\ccsdesc[500]{Human-centered computing~Empirical studies in HCI}

\keywords{algorithms, affective polarization, partisan animosity, social media AIs, social media users}

\begin{teaserfigure}
  \includegraphics[width=\textwidth]{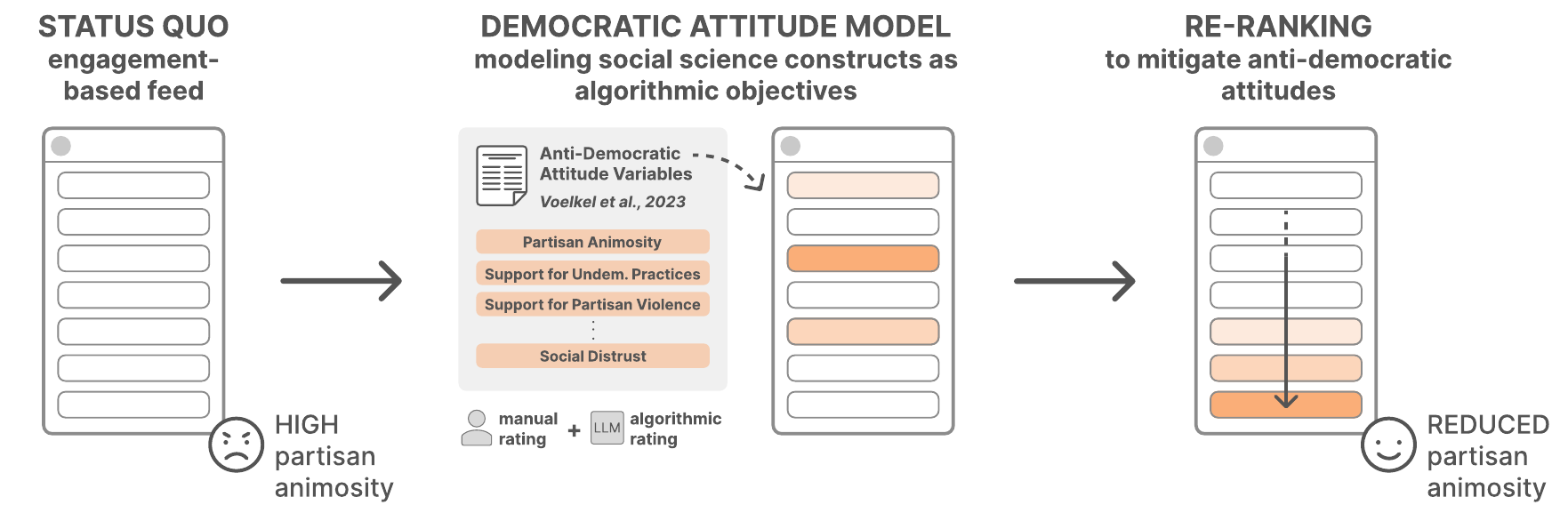}
  \caption{
  (Left) Social media tends to orient objective functions around observable variables like user engagement, which may drive partisan animosity.
  (Center) Our \textit{societal objective function} method models social scientific constructs as algorithmic objectives. We translate anti-democratic attitude variables to qualitative codebooks for manual ratings and LLM-based algorithmic ratings that produce an anti-democratic attitude model for political social media posts.
  (Right) We re-rank social media feeds to mitigate anti-democratic attitudes and observe reductions in partisan animosity.
  }
  \label{fig:pull_figure}
\end{teaserfigure}

\maketitle

\section{Introduction}
Can social media support a healthy democracy? Social media AIs such as feed ranking algorithms bear nontrivial responsibility in how people perceive contentious political issues \cite{huszar2022algorithmic}. Researchers have long been interested in the effects of social media AIs on partisan information consumption (e.g.,~\cite{bail2022breaking, guess2019accurate, robertson2023users, bakshy2015exposure}) because these AIs shape people's beliefs \cite{brady2023overperception}, affect their well-being \cite{hancock2022psychological, kreski2021social}, and change their behaviors \cite{vannucci2020social}.

A key outcome of interest in a healthy democracy is \textit{partisan animosity}~\cite{milli2023twitter, tornberg2022digital}: negative thoughts, feelings and behaviours towards a political out-group. In the United States, for example, over 80\% of members of both political parties express concern that the country is growing increasingly divided~\cite{pewanimosity}. High levels of partisan animosity undermine our collective will to address public issues~\cite{hartman2022interventions}. Unfortunately, mounting evidence suggests that social media is associated with increases in partisan animosity in established democracies~\cite{lorenz2023systematic, milli2023twitter, tornberg2022digital}. While this effect can be mitigated by users' own behavior and choices~\cite{robertson2023users,bakshy2015exposure}, many studies demonstrate that feed algorithms can amplify political content~\cite{huszar2022algorithmic} and decrease people's trust in democracy~\cite{lorenz2023systematic}.

Could we design social media AIs to more directly consider their impact on partisan animosity?
Social media AIs, and specifically feed ranking algorithms, are centered around accurately predicting engagement signals such as likes and clicks~\cite{narayanan2023,huszar2022algorithmic,liu2010personalized}.
In contrast, partisan animosity may not be detectable with engagement behavior alone, and it can even be anti-correlated with traditional engagement signals: maximizing engagement can amplify anti-social behaviors~\cite{are2020instagram, munn2020angry} and focus attention on pro-attitudinal political content~\cite{sunstein2001http,rowland2011filter}. Lacking a method for modeling partisan animosity directly in the algorithm, researchers and practitioners instead engage in mitigation strategies such as ideologically balanced feeds~\cite{celis2019controlling} and content moderation policies~\cite{gillespie2018custodians}. However, these mitigations are indirect: they are not \textit{designed} to treat a societal value (here, reducing partisan animosity) as a first-class algorithmic objective.

\begin{figure}[!t]
    \centering
    \includegraphics[width=1 \textwidth]{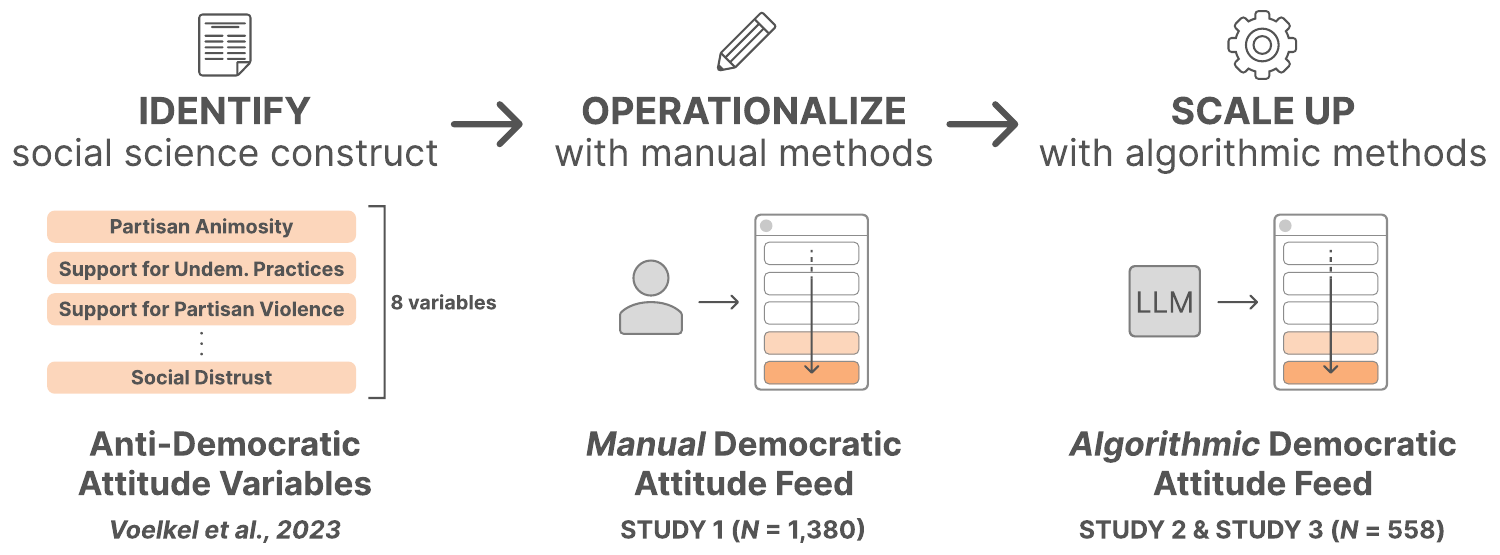}
    \caption{Steps of our \textit{societal objective function} method: (1) Identify a well-established social scientific construct, such as the construct of anti-democratic attitudes; (2) Operationalize the construct with manual rating methods, as shown with our manual democratic attitude feed in Study 1; (3) Scale up the ratings with algorithmic methods using an LLM, as shown with our algorithmic democratic attitude feed in Studies 2 \& 3.}
    \label{fig:method}
\end{figure}

To address this gap, we demonstrate a method for integrating established and vetted social scientific constructs into objective functions, which we term \textit{societal objective functions}. Specifically, we observe that survey measurements and qualitative codebooks in the social and behavioral sciences have long needed to be precise to facilitate reliability, and recent research has found that this precision enables their translation into prompts interpretable by large language models (LLMs) such as GPT-4 ~\cite{xiao2023qualitative, ziems2023large, huang2023chatgptAnnotation, wang2021reducelabeling, do2022augmentedSocialScientist}. In this work, we go one step further to embed algorithms with technical objectives derived from these social science constructs. In this case, we draw on the political science literature and specifically its measurement of \textit{anti-democratic attitudes}. This measure, which was recently tested in a large study that received widespread attention~\cite{voelkel2023megastudy}, spans eight variables that describe willingness to engage in good faith in the democratic process: partisan animosity, support for undemocratic practices, support for partisan violence, support for undemocratic candidates, opposition to bipartisanship, social distrust, social distance, and biased evaluation of politicized facts. We adapt each of these eight variables into prompts for a large language model and then combine them into a joint \textit{democratic attitude model} that agrees with manual human annotation on political social media posts. This democratic attitude model can be applied at scale to estimate the impact of every post in a social media feed on anti-democratic attitudes, enabling democratic attitudes to be integrated into a social media feed ranking algorithm. We summarize the steps of our societal objective function method in Figure~\ref{fig:method} and several of the key terms used throughout the paper in Table~\ref{tab:term_bank}.

\begin{table*}[!tb]
\centering
\footnotesize
\begin{tabular}{p{0.22\textwidth} p{0.78\textwidth}}
\toprule
\textbf{Term} & \textbf{Definition} \\
\midrule
\textit{Partisan animosity} & {Negative thoughts,
feelings and behaviours towards a political out-group.}\\[0.1cm]
\textit{Anti-democratic attitudes} & {Eight variables from the political science literature, combined into a measure in ~\citet{voelkel2023megastudy} to describe willingness to engage in good faith in the democratic process: (1)~partisan animosity, (2)~support for undemocratic practices, (3)~support for partisan violence, (4)~support for undemocratic candidates, (5)~opposition to bipartisanship, (6)~social distrust, (7)~social distance, and (8)~biased evaluation of politicized facts.}\\[0.1cm]
\textit{Democratic attitude model} & {Our AI model that leverages zero-shot prompts to a large language model to rate the impact of social media posts on anti-democratic attitudes. The model takes social media posts as input and generates scores for each of the eight anti-democratic attitudes. These scores estimate the extent to which the post promotes each of the eight anti-democratic attitudes, and are summed to a single democratic attitude score.}\\[0.1cm]
\textit{Democratic attitude feed} & {A re-ranked social media feed that uses the democratic attitude model to perform downranking, generate content warnings, or perform content removal-and-replacement.}\\[0.1cm]
\textit{Societal objective function} & {Our method that integrates established and vetted social scientific constructs into objective functions by translating from constructs to manual codebooks to algorithmic ranking using zero-shot prompting with LLMs.}\\[0.1cm]
\bottomrule
\end{tabular}
\vspace{-0.01in}
\caption{A summary of key terms used throughout the paper.}
\label{tab:term_bank}
\end{table*}

\begin{figure}[!t]
    \centering
    \includegraphics[width=1 \textwidth]{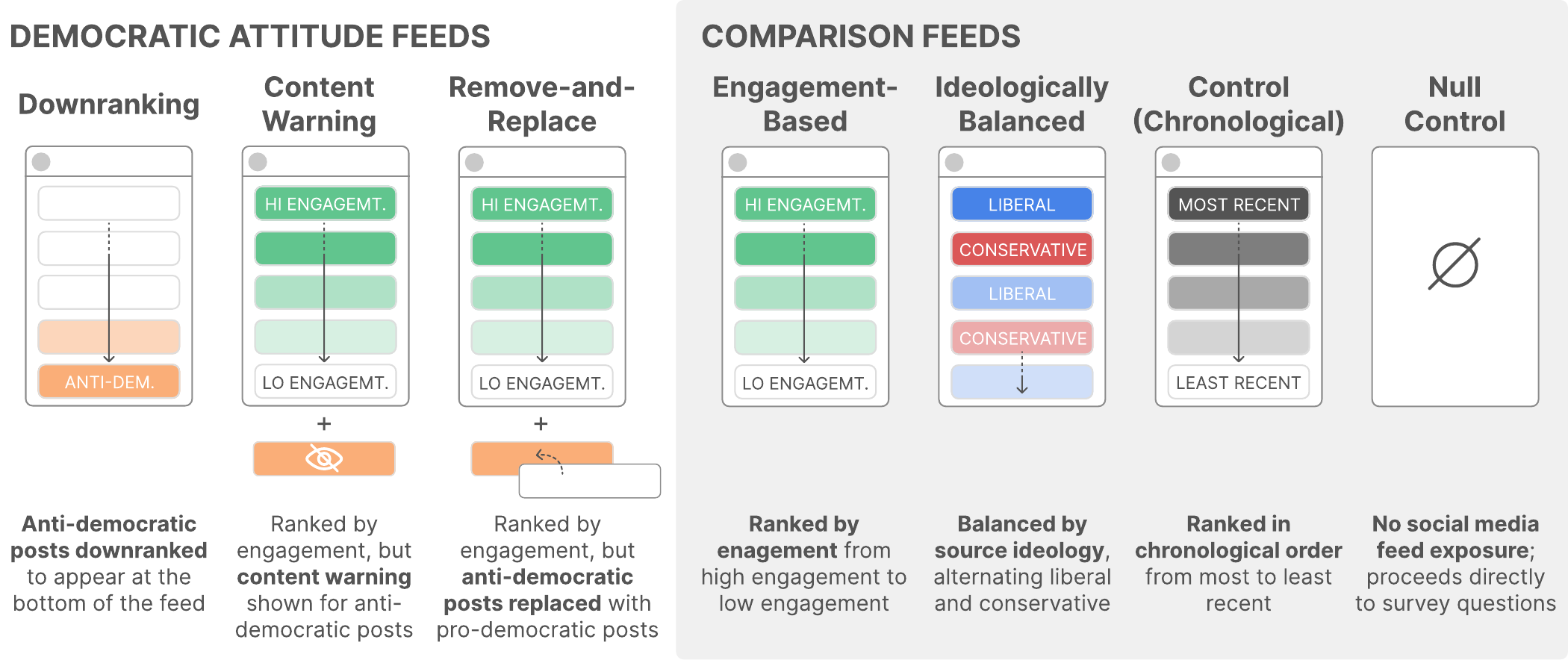}
    \caption{Summary of our seven feed ranking conditions. The \textit{democratic attitude feeds} incorporate our anti-democratic attitude model with either: (1) Downranking, (2) Content Warning, or (3) Remove-and-Replace feeds. The \textit{comparison feeds} capture a range of existing feed ranking methods including (4) Engagement-Based, (5) Ideologically Balanced, or (6) Chronological feeds as well as (7) a Null Control where no feed is shown.}
    \label{fig:condition}
\end{figure}

We integrated this democratic attitude model into a feed ranking algorithm and tested it across three studies to investigate its impact on partisan animosity as well as traditional platform engagement measures. In Study 1, we began with \textit{manual, human labels} for the construct to establish its behavioral effectiveness before translating it into an AI: we manually annotated a prepared feed of political social media content using the existing anti-democratic attitude scale (Krippendorff's $\alpha = .895$). We conduct a preregistered, between-subjects online experiment among U.S. partisans (Democrats or Republicans, $N$ = 1,380) to examine the effect of \textit{manual democratic attitude feeds} that use these manual labels for re-ranking, removal, and warning against a traditional \textit{engagement-based feed} or a \textit{chronological feed} on participants' partisan animosity, support for undemocratic practices, feed-level satisfaction metrics, and engagement, as shown in Figure~\ref{fig:condition}.

We found that our democratic attitude feeds---specifically the \textit{downranking feed} and \textit{remove-and-replace feed}---significantly reduced partisan animosity compared to a traditional engagement-based feed. These results hold for both conservatives and liberals, and hold without compromising participants' engagement level or their ratings of their experience on the platform. Though freedom of speech is cited as a frequent concern for algorithmic feed ranking methods, we found that the democratic attitude feeds that re-rank or remove do not prompt perceived threats to freedom of speech. 

Next, we translated from manual annotation to an automated societal objective function. In Study 2, we used zero-shot prompting with a large language model (LLM) to translate the original anti-democratic attitude construct into a democratic attitude model via GPT-4. Testing the accuracy of the democratic attitude model, we observed that the democratic attitude model ratings correlated highly with those from human coders (Spearman's $\rho = .75$), indicating that the manual approach can be replicated at scale using LLMs. Then, in Study 3, we replicated Study 1, this time using the algorithmic democratic attitude model rather than the manual labels. The original results replicated; the algorithmic democratic attitude feed significantly reduced partisan animosity with an effect size similar to that of the manual democratic attitude feed.

Our work introduces a novel method of translating social science theory to embed societal values in feeds via algorithmic objectives, which we term \textit{societal objective functions}. We accomplish this by operationalizing social science constructs into a manual codebook, using the codebook to manually re-rank feeds, and validating the effect that re-ranked feeds have on societal outcomes with online experiments (Study 1). Then, we scale up the codebook using zero-shot prompting with LLMs and show that algorithmic ranking can replicate both manual feed ranking (Study 2) and online experiment results (Study 3). 

We demonstrate the viability of our method with the specific construct of democratic attitudes, but our approach carries implications for how we might incorporate a much broader set of societal values into social media feeds. We note that today's social media \textit{already} encodes values, but these values are typically implicit and focused on individual user outcomes, such as engagement or time spent on the platform~\cite{bernstein2023embeddingJOTS}. Our work aims to make these values more explicit and tuneable. Toward this goal, in this paper we develop an existence proof via one value, pro-democratic attitudes, to demonstrate the feasibility of the method and how it might apply to a broader plurality of values. A plural set of values are ultimately important to capture for social media feed ranking, and may be amenable to bottom-up, participatory processes. Given a new societal value of interest (e.g., wellbeing, cultural diversity, environmental sustainability), future work might leverage this method to identify relevant social science constructs and instantiate them into additional algorithmic objectives to embed societal values in feeds. Today's social media feeds lack an understanding of the impact they may have on societal values like mitigating anti-democratic attitudes---our work validates a promising pathway to directly embed societal values into the objectives that drive social media AIs.

\section{Related Work}
Algorithmic social media feed ranking has consequences not just for individual users, but also for society. Social media AIs shape people's beliefs \cite{brady2023overperception}, affect their mental well-being \cite{hancock2022psychological, kreski2021social}, and change their behaviors \cite{vannucci2020social}. These consequences accrue to the individual, of course, but also aggregate to the societal level, for example through their impact on democratic discourse. 

\subsection{Encoding Societal Values into Social Media AIs }

Trading off societal outcomes is a complex issue. Today, most recommender systems, including those that power social media feeds, center around individual user experiences. Recommendation algorithms rely on key metrics, or objectives, that determine how to score candidate items to filter, rank, and display to users~\cite{eckles_2022}. 
Most commonly, feed ranking systems focus on metrics of user engagement (e.g., clicks, views, comments, likes), which serve as proxies for user satisfaction and, ultimately, platform revenue~\cite{milli2021optimizing,ciampaglia2018algorithmic}. These algorithms can end up maximizing individual experience at the cost of societal values such as pro-democratic attitudes or partisan animosity. For instance, maximizing engagement can amplify anti-social behaviors such as online harassment \cite{are2020instagram, munn2020angry} or focus our attention on pro-attitudinal political content \cite{sunstein2001http,rowland2011filter}. 

Though engagement metrics can be misleading as proxies for user benefit~\cite{kleinberg2022challenge}, these metrics are amenable to measurement and modeling since behavioral traces from organic platform use are the most abundant form of data (in contrast to explicit signals like user surveys or human annotation)~\cite{stray2022building}. Some platforms periodically survey their users to gather high-level feedback, such as Facebook's survey that asked users to gauge whether posts were good or bad for the world~\cite{fb_goodfortheworld}, but survey feedback is much more limited in scale, so it is challenging to formulate algorithmic objectives around this feedback. 

To address these issues, social media platforms have explored efforts that incorporate some notions of societal values into ranking models. Content moderation models are a common strategy to ensure that content does not violate platform policies or community guidelines~\cite{gillespie2018custodians}, and such models are often developed with the help of crowdsourced data annotations~\cite{youtube_four_rs}. Beyond content moderation, platforms have also taken measures to combat the societal harms posed by content such as terrorism or extremism, misinformation, and violence with automated detection and removal~\cite{fb_enforcement, twitter_enforcement, youtube_enforcement}. On the modeling side, recent developments in AI value alignment, such as Constitutional AI~\cite{bai2022constitutional} and reinforcement learning from human feedback (RLHF)~\cite{ouyang2022training} present technical methods for holistically steering AI models to better align with human descriptions and demonstrations of desired behavior.
These steps are critical in handling acute, negative societal values and preserving general norms of communication. Our work seeks to extend these approaches to not only combat stark policy violations, but also to capture a wider range of societal values that goes beyond policy enforcement.

Meanwhile, researchers have also investigated interventions on social media feeds to counter various user harms that may relate to societal values. For example, Gobo addressed the lack of control over social media feeds by introducing a system that allowed users to aggregate and filter content across platforms~\cite{bhargava2019gobo}. In a related vein, the HabitLab system sought to grant users agency over social media usage by enacting productivity interventions~\cite{kovacs2018habitlab}. Other work has explored interventions that expose users to alternative feeds that may productively differ from those that they typically see: ``Blue Feed, Red Feed'' presented contrasting liberal and conservative Facebook feeds~\cite{bluefeed_redfeed}, and researchers have explored feed re-ranking interventions aimed at achieving ideological balance~\cite{celis2019controlling} or bridging-based ranking to build trust across divides~\cite{ovadya2023bridging}. However, some researchers argue that those "more balanced" feed designs may not be effective enough to reduce partisan animosity \cite{nelimarkka2018social}. Therefore, we build on this line of prior work in presenting concrete implementation strategies to intervene on social media feeds more effectively. While prior approaches may not entirely capture or align with societal values, we address the challenge of bringing implementation and societal values in line.

Given the difficulty of quantifying societal values and the resulting scarcity of social media ranking models that optimize for such values, we need new research agendas that might address these issues. 
Recent work has started to map and articulate the space of human values that could be instantiated in social media.
For example,~\citet{stray2022building} provide a broad set of over 30 human values compiled from multidisciplinary experts, such as well-being, freedom of expression, and civic engagement. 
Prior work has introduced valuable process frameworks that outline how to bridge from human values to algorithmic systems~\cite{stray2022building, zhu2018valuesensitivealgorithm}, but it remains challenging to connect those values to implementation. Our work thus sets out to connect from societal values to social science constructs to concrete algorithmic implementations, and we demonstrate the promise of this approach with a societal value of democratic attitudes.

\subsection{Social Media Algorithms and Partisan Animosity}

In light of a growing body of literature suggesting harmful connections between social media algorithms and democracy, we focus on democratic attitudes in our work. Partisan animosity refers to tendency of partisans to hold negative views of opposing partisans, but positive views of co-partisans~\cite{iyengar2015polarization,iyengar2019origins}. 
Social scientists, practitioners, and activists have long been interested in reducing partisan animosity among Americans \cite{wojcieszak2020can, ahler2018parties}.
In the United States, this divide appears to be growing more extreme~\cite{boxell2021trends}, driving worry about undemocratic practices and existential risks to democracy \cite{kingzette2021affective}. 
Partisan animosity is often associated with affective polarization \cite{voelkel2023megastudy}; in this study, we focus on affective polarization instead of issue polarization, which reflects partisans' disagreement about certain political issues (e.g., abortion, gun control) \cite{hartman2022interventions}. Some researchers operationalize \textit{affective polarization} by measuring disliking partisanship in general instead of a specific party \cite{klar2018affective}, whereas others introduce new terms or measurements such as ``partyism'' to describe the hostility and aversion to a certain political party \cite{sunstein2015partyism}. Given the diversity and disparity of operationalizations of affective polarization in past research, we choose to focus on a broader term of \textit{partisan animosity} following \citet{hartman2022interventions}. As in this prior work, we define partisan animosity as hostility and aversion to the opposing party, and measure this with ratings of warmth on a feeling thermometer (e.g., \cite{iyengar2019origins, voelkel2023megastudy}).

Since digital media introduces heavy personalization and amplifies messages at a vastly different speed and scale than prior forms of media, researchers have investigated potential links between algorithmic behavior, media consumption, and user polarization.  For example, there is evidence that the Twitter algorithm amplifies content from the political right~\cite{huszar2022algorithmic} and that the ranking algorithm amplifies partisanship and out-group animosity, especially for political tweets~\cite{milli2023twitter}. Studies have similarly indicated that Facebook usage promotes political polarization~\cite{alcott2020welfare}.
However, there is disagreement about the mechanisms by which social media influences polarization. Earlier research hypothesized about the role of echo chambers (or selective exposure) in isolating individuals or communities into homogeneous clusters and driving them towards more extreme and divergent positions \cite{sunstein2001http}. More recent evidence supports alternative mechanisms like partisan sorting, whereby polarization is not driven by isolation, but by repeated \textit{exposure} to individuals outside of local networks, which might cause local conflicts to align on global partisan lines~\cite{tornberg2022digital}.  

Thus, there are a variety of possible mechanisms by which algorithmic ranking on social media ultimately influences the political beliefs of users. While much of this prior work studies the impact of existing social media (e.g., \cite{gonzalez2023asymmetric}), some recent studies suggest a few alternative design options to combat polarization on social media such as diversifying information sources \cite{nelimarkka2019re}, highlighting agreeable items via browser widgets \cite{munson2010presenting, munson2013encouraging}, and designing political deliberation environments \cite{semaan2015designing}. Our work aims to build on these prior findings to \textit{redesign} social media ranking algorithms and experimentally tie these design decisions to users’ political beliefs. Most prior work either uses observational data to investigate whether current social media AIs exacerbate partisan animosity \cite{milli2023twitter, tornberg2022digital} or examines the effect of bottom-up interventions on participants' anti-democratic attitudes such as reducing exposure to content from like-minded sources \cite{nyhan2023like} and reshared content \cite{guess2023reshares}. To our knowledge, prior work has not taken a top-down approach to implement high-level democratic values into feed algorithms. Even though some prior work measured the influence of people's social media consumption on political activities such as voting decisions \cite{maruyama2014hybrid}, we believe no prior work has taken a preemptive step to embed social science measurement in the construction of one's social media feed. Adding to prior literature, our work translates social science measures of anti-democratic attitudes directly into objective functions and examines effects on partisan animosity.

\subsection{Democratic Values As A Lever: A Sociotechnical Approach}
Given the potential harms that current engagement-based social media feeds may pose to democracy, our work uses democratic values as a lever to examine whether a feed that embeds democratic values could reduce partisan animosity. The key question then becomes: how do we operationalize democratic values? 
To address this question, we first adopt a measure of anti-democratic attitudes from political science research \cite{voelkel2023megastudy, hartman2022interventions}, where the construct has been previously vetted and tested. We utilize its eight sub-scales to label each social media post, producing a continuous rating of the extent to which each post potentially impacts anti-democratic attitudes, and then replicate the manual rating using an algorithmic approach. 

We do not claim that the construct of anti-democratic attitudes is the only construct that might matter---far from it---nor that its current operationalization is perfect. However, we find it far more productive to draw on social science expertise rather than re-invent the wheel. We chose to operationalize anti-democratic attitudes as an example due to its recent large-scale vetting in a large study by ~\citet{voelkel2023megastudy} that includes 25 interventions ($N$ = 32,059). In this study, anti-democratic attitudes are measured for US partisans using eight variables, namely \textit{partisan animosity}, \textit{support for undemocratic practices}, \textit{support for partisan violence}, \textit{support for undemocratic candidates}, \textit{opposition to bipartisanship}, \textit{social distrust}, \textit{social distance}, \textit{biased evaluation of politicized facts}. 

Related work by \citet{hartman2022interventions} defined \textit{partisan animosity} as negative thoughts, feelings or behaviors towards outgroup; they argue that partisan animosity is an umbrella term that synthesizes a variety of concepts such as affective polarization, interpersonal polarization, and political sectarianism \cite{hartman2022interventions}. The Voelkel et al. megastudy found that almost all of their interventions successfully reduced partisan animosity, and several interventions reduced support for undemocratic practices and partisan violence \cite{voelkel2023megastudy}. Other work in the field of political science also measured a subset of these eight outcome variables. For instance, \citet{hartman2022interventions} measured partisan animosity because they argue that the rising partisan animosity is associated with the decrease in support for democracy. \citet{druckman2023correcting} measured partisan animosity, support for undemocratic practices, and support for partisan violence through a survey experiment, and found that correcting misperceptions of out-partisans can decrease American legislators’ support for undemocratic practices and marginally significantly reduce their partisan animosity.  

We envision that future work can develop a larger set of these constructs to integrate into social media AIs and trade off amongst one another. One reason we choose anti-democratic attitudes as our construct of interest here is that it is relatively broad, measuring not just partisan animosity but also several other subscales, which aligns with realistic social media settings where there are multiple competing values to consider and trade off. The Voelkel et al. study was also conducted by a nationally-representative pool of social science researchers and studied a large, diverse sample of participants, so it represents a significant contribution to the literature on anti-democratic attitudes. However, our approach is not restricted to this particular paper and choice of construct, and the same approach could be used to translate other constructs to algorithmic objectives.

\section{Societal Objective Functions}

\subsection{Operationalizing Anti-Democratic Attitudes into Social Media AIs}

We introduce the term \textit{societal objective function} to refer to our method of translating well-established social science constructs into an AI objective function.
In this work, the goal of our algorithmic objective is to reduce partisan animosity through social media feeds. 

\subsubsection{The anti-democratic attitudes construct}

Creating a societal objective function begins with anchoring to a construct in the social and behavioral sciences. In our case, as mentioned, we use anti-democratic attitudes. Detailed definitions and example measures of each of the eight variables are shown in Table ~\ref{tab:megastudy}. These variables were selected because they are important measurements related to psychology underlying polarization and democracy \cite{voelkel2023megastudy}.

\begin{table}[]
\resizebox{\textwidth}{!}{%
\begin{tabular}{@{}llllll@{}}
\toprule
\multicolumn{1}{c}{\textbf{Outcome Variables}} & \multicolumn{1}{c}{\textbf{Definition}}                                                                             & \multicolumn{2}{c}{\textbf{Example Item}}                                                                                                                                                                                                         \\ \midrule
Partisan Animosity                             & Dislike for opposing partisans                                                                                       & \multicolumn{2}{l}{How would you rate {[}Democrats/Republicans{]}?}                                                                                                                                                                                  \\ \midrule
Support for Undemocratic Practices             & \begin{tabular}[c]{@{}l@{}}Willingness to forgo democratic \\ principles for partisan gain\end{tabular}             & \multicolumn{2}{l}{\begin{tabular}[c]{@{}l@{}}{[}Republicans/Democrats{]} should not accept\\ the results of elections if they lose.\end{tabular}}                                                                                                   \\ \midrule
Support for Partisan Violence                  & \begin{tabular}[c]{@{}l@{}}Willingness to use violent tactics \\ against outpartisans\end{tabular}                  & \multicolumn{2}{l}{\begin{tabular}[c]{@{}l@{}}How much do you feel it is justified for \\ {[}Republicans/Democrats{]} to use violence if the \\ {[}Democratic/Republican{]} party wins more races \\ in the next election?\end{tabular}}             \\ \midrule
Support for Undemocratic Candidates            & \begin{tabular}[c]{@{}l@{}}Willingness to ignore undemocratic \\ practices to elect inparty candidates\end{tabular} & \multicolumn{2}{l}{\begin{tabular}[c]{@{}l@{}}How would you vote if you learned that the \\ {[}Republican / Democratic{]} candidate said that \\ {[}Republicans /Democrats{]} should not accept the \\ results of elections they lose?\end{tabular}} \\ \midrule
Opposition to Bipartisan Cooperation           & \begin{tabular}[c]{@{}l@{}}Resistance to cross-partisan \\ collaboration\end{tabular}                               & \multicolumn{2}{l}{\begin{tabular}[c]{@{}l@{}}Generally speaking, would you say that most \\ people can be trusted, or that you need to be \\ very careful in dealing with people?\end{tabular}}                                                     \\ \midrule
Social Distrust                                & Distrust of people in general                                                                                       & \multicolumn{2}{l}{\begin{tabular}[c]{@{}l@{}}How comfortable are you having close personal \\ friends who are {[}Democrats/Republicans{]}?\end{tabular}}                                                                                            \\ \midrule
Social Distance                                & \begin{tabular}[c]{@{}l@{}}Resistance to interpersonal contact \\ with outpartisans\end{tabular}                    & \multicolumn{2}{l}{\begin{tabular}[c]{@{}l@{}}To what extent would you like to see Democratic \\ and Republican elected representatives work together?\end{tabular}}                                                                                 \\ \midrule
Biased Evaluation of Politicized Facts         & \begin{tabular}[c]{@{}l@{}}Skepticism of facts that favor the \\ worldview of the other party\end{tabular}          & \multicolumn{2}{l}{\begin{tabular}[c]{@{}l@{}}{[}Joe Biden/Donald Trump{]} was lawfully elected \\ President in the {[}2020/2016{]} election against \\ {[}Donald Trump/Hillary Clinton{]}.\end{tabular}}                                            \\ \bottomrule
\end{tabular}}
\vspace{0.05in}
\caption{Definitions and example measurement of eight anti-democratic attitude outcome variables in the political science literature~\cite{voelkel2023megastudy}.}
\label{tab:megastudy}
\end{table}

\subsubsection{Dataset}
To generate the social media feed stimulus for our three studies, we source real-world posts drawn from CrowdTangle, a tool hosted by Meta that allows external parties to monitor public posts on Facebook \cite{fletcher2018measuring}. Given our focus on democratic values, we source political posts on CrowdTangle by filtering to several politics-related page categories\footnote{Page categories: political organization, political candidate, political ideology, political party, politics, politician, government organization, government official, public services government} and select those with the highest number of total interactions (likes, shares, comments, and reactions). Using this method, we gather $10,000$ political Facebook posts from CrowdTangle that were posted between January 1 and February 1, 2023. We then use the systematic random sampling method to select a final inventory of $n$ = 405 posts for manual labeling. Specifically, we split the CrowdTangle posts into buckets based on overall weighted engagement counts (e.g., Likes), and sample posts across these buckets to ensure a broad range of engagement in the posts in our study.

\subsubsection{Translating from manual coding to feed ranking interfaces}

Then, two members of the research team served as expert annotators to rate each political post using the anti-democratic attitude scale. For each anti-democratic outcome variable, a rating is given on a 3-point scale, with a score of 3 indicating strong presence of that variable (e.g., strong partisan animosity), a score of 2 indicating some presence (e.g., some partisan animosity), and a score of 1 indicating no presence (e.g., no evidence of partisan animosity). The two researchers adapted the existing construct from the literature deductively into a detailed coding scheme for each of the eight variables. Notably, the \citet{voelkel2023megastudy} study evaluates anti-democratic attitudes of \textit{people}, while we seek to evaluate \textit{social media posts}. Thus, we choose to rate the posts by the extent to which they would promote a given anti-democratic attitude in the audience. Some new factors of the constructs, such as ``emotion or exaggeration,'' also emerged inductively in the process of creating the coding scheme. For instance, for partisan animosity, they tagged two factors in their rating procedure---factor A: partisan name-calling and factor B: emotion or exaggeration. 

In this case, a post was given a rating of 1 if neither factor applied, a rating of 2 if only one of the factors applied, and a rating of 3 if both factors applied. 

While our operationalization is anchored in the prior literature, we anticipate that future work will offer further refinements of the codebook --- our work is not dependent on the exact operationalization.

Each post is given a total of 8 ratings, one for each of the variables, which are then summed to arrive at a total \textit{democratic attitude} score (min = 8, max = 24).
The two independent coders achieved strong inter-coder reliability (Krippendorff's $\alpha$ = .895) and used the above process to code all 405 political Facebook posts in the inventory. Full inter-coder reliability results are shown in Table~\ref{tab:manual_irr}.

The manual democratic attitude scores are then used to re-rank social media feed interfaces. In Study 1, we use this approach to compare democratic attitude feeds with status quo feed ranking methods. For example, we sort our inventory of social media posts to produce feeds ranked by manually-rated democratic attitude scores or by total interaction scores from CrowdTangle. The total interaction score is a weighted sum of the following engagement metrics: Share, Comment, Like, Love, Wow, Haha, Sad, Angry, and Care. Each engagement metric is given equal weight. Our dataset includes posts with total interaction scores ranging from 24 (low interaction) to 92,520 (high interaction). Our full set of feed ranking conditions is described in Section~\ref{section:feed_ranking_conditions}.

\subsubsection{Using LLMs to replicate manual coding at scale} 

The final step of creating a societal objective function involves using LLMs to replicate the same social science construct at scale. In our work, we scale up the anti-democratic attitude construct by turning the eight variables into zero-shot classification prompts for a large language model. These prompts are built on the exact same coding scheme developed for the manual raters, and used as inputs to the GPT-4 large language model to output ratings. Specifically, we prompt GPT-4 to rate each social media post from our social media post inventory dataset (development set: $n$ = 205; test set: $n$ = 200) on all eight variables. The development set is used to refine and iterate on the prompts, and the test set is set aside to conduct final performance evaluations. The full prompts are included in Appendix~\ref{appendix:prompts}. Then, as with the manual rating procedure, we sum the eight anti-democratic attitude scores to produce the total anti-democratic attitude score. 

In Study 2, we first compare GPT-4's ratings with manual ratings. Then, in Study 3, we use ratings from GPT-4 to produce an LLM-ranked feed condition to compare against Study 1's manually-ranked and engagement-ranked feed conditions.

\subsection{Feed Ranking Conditions and Hypotheses}
 
We pre-registered our research questions and hypotheses on Open Science Framework (OSF) \cite{foster2017open} and conduct two online experiments (Study 1: $N$ = 1,380; Study 3: $N$ = 558) to measure the impact of democratic attitude feeds on US partisans' partisan animosity, support for undemocratic practices, feed-level satisfaction, and engagement.

\subsubsection{Feed ranking conditions}
\label{section:feed_ranking_conditions}

Past work has examined different approaches to reduce the societal harm caused by social media, such as using content moderation~\cite{kozyreva2023resolving}, downranking or removing harmful content or misinformation~\cite{epstein2020will}, placing content warnings to warn potential viewers that content is sensitive or may bring up difficult emotions~\cite{haimson2020trans}, or displaying more balanced information in one's social media feed (e.g., ideologically balanced content from both liberal and conservative sources \cite{celis2019controlling}). 

Building on prior work, we propose three democratic attitude feeds, including:
\begin{enumerate}
    \item \textit{Downranking} feed (i.e., ranked by anti-democratic attitude score such that posts with stronger anti-democratic attitudes are placed at the bottom of the feed)

    \item \textit{Content Warning} feed which mirrors designs commonly used by real-world social media platforms to mask harmful content (i.e., ranked by engagement, but anti-democratic posts are blurred out with content warnings)
    \item \textit{Remove-and-Replace} feed (i.e., ranked by engagement, but anti-democratic posts are replaced with pro-democratic posts sourced from our dataset inventory ($n$ = 405))
\end{enumerate}

Current feed ranking systems often focus on optimizing users' engagement to increase user retention \cite{wu2017returning} and maximize profit \cite{ciampaglia2018algorithmic}, which is likely to up-rank the most controversial content and thus increase partisan animosity. Many scholars have articulated their concerns about engagement-based feeds and often compare the impact of engagement-based feeds with reverse-chronologically ordered feeds that are free from algorithmic curation \cite{paek2010predicting, huszar2022algorithmic}. We thus compare our democratic attitude feeds against both engagement-based feeds that emulate the status quo of feed ranking and reverse-chronological feeds that serve as a control condition. In addition, we compare our democratic attitude feeds against the ideologically balanced approaches proposed by prior work \cite{celis2019controlling}. Then, to estimate the baseline level of partisan animosity, we include a null control condition where participants are not exposed to any social media feed. We thus arrive at four comparison conditions: (1)~an \textit{Engagement-Based} feed, (2)~an \textit{Ideologically Balanced} feed, (3)~a \textit{Control (Chronological)} feed, and (4)~a \textit{Null Control} condition with no feed shown. See Figure~\ref{fig:condition} for a detailed description of all feed ranking conditions.

\subsubsection{Research Questions and Hypotheses}

Our study intends to examine the following overarching research questions: 

\begin{enumerate}
    \item[\textbf{RQ1:}] How will partisans’ partisan animosity and support for undemocratic practices differ after exposure to different social media feeds?

    \item[\textbf{RQ2:}] How will partisans’ level of satisfaction and engagement with feeds differ after exposure to different social media feeds?
\end{enumerate}

Prior work found that current social media platforms' ranking algorithms amplify political content \cite{huszar2022algorithmic} and increase opinion polarization \cite{ciampaglia2018algorithmic, chitra2020analyzing, rowland2011filter}. There is a growing literature on the ways in which ranking algorithms trained on engagement data might facilitate the formation of ``echo chambers'' \cite{sunstein2001http} or ``filter bubbles'' \cite{rowland2011filter}. Recent work found that interventions such as changing public discourse and transforming political structures or correcting misconceptions and highlighting commonalities can decrease people's animosity \cite{voelkel2023megastudy, hartman2022interventions}. Following a large body of literature on partisan animosity, we also predict that our democratic attitude feed can effectively reduce partisan animosity. Thus, we predict \textbf{H1}:

\begin{enumerate}
    \item[\textbf{H1:}] Partisans exposed to the (a) downranking, (b) content warning, and (c) removal feed conditions on social media will reduce partisan animosity compared to partisans in the engagement feed and chronological feed (controls).
\end{enumerate}

\citet{voelkel2023megastudy} point out that many scholars are also concerned about Americans' support for undemocratic practices besides the level of dislike between rival partisans (i.e., partisan animosity) \cite{finkel2020political}. Given the increasing scholarly attention to the importance of support for undemocratic practices, we also predict \textbf{H2}:

\begin{enumerate}
    \item[\textbf{H2:}] Partisans exposed to the (a) downranking, (b) content warning, and (c) removal feed conditions on social media will reduce support for undemocratic practices compared to partisans in the engagement feed and chronological feed (controls).
\end{enumerate}

Recent prior work found that people's perceptions of content moderation depends on partisanship: Republicans were more likely to oppose content moderation than Democrats possibly because they consider it a threat to freedom of speech \cite{kozyreva2023resolving}. Other studies also found that content moderation may trigger people's concerns about platform censorship \cite{riedl2022antecedents}. Based on prior work, we predict that partisans exposed to the content warning feed will perceive a higher level of threat to freedom of speech compared to other feeds and propose \textbf{H3}.

\begin{enumerate}
    \item[\textbf{H3:}] Partisans exposed to the content warning feed will perceive a higher level of threat to freedom of speech compared to partisans exposed to other feeds.
\end{enumerate}
\section{Study 1: The Impact of Manually Labeled Democratic Attitude Feeds}

In Study 1, we examined the impact of manually-generated democratic attitude feeds on the partisan animosity of US partisans. By experimenting with a manual version of our ranking algorithm, we aim to understand the effect of democratic attitude feeds given a hypothetical ``perfect'' AI that reflects the initial social science construct via expert annotation. 

\subsection{Method}
\subsubsection{Experimental Design}
We created a social media feed named \textit{PolitiFeed} that consists of a wide variety of real-world political posts, ranging from posts with high anti-democratic attitudes and low anti-democratic attitudes. We conducted a between-subjects design among US partisans ($N$ = 1,380) in March 2023 with seven conditions, respectively the downranking ($n$ = 191), content warning ($n$ = 192), removal ($n$ = 193), engagement-based ($n$ = 207), ideologically balanced ($n$ = 197), chronological feed ($n$ = 198), and null control ($n$ = 202). We randomly assigned participants to one of the seven conditions. In different treatment conditions, participants were exposed to scrollable social media feeds with the same inventory content but different ranking methods. In the control condition, participants were exposed to a chronological social media feed. Participants in each feed condition were asked to read 60 political posts sourced from the inventory dataset, as 60 posts roughly replicate two full loads of the Twitter timeline. In the null control condition, participants were not exposed to any feeds. Participants spent on average 414.51 seconds on viewing the feed and 631.31 seconds on answering the questionnaire. 

\subsubsection{Participants} 
To detect a main effect of condition and interactions with people's partisan affiliation, a priori power analysis using G*Power determined that a total sample of at least 1,369 participants would be needed to achieve 80\% power if $\alpha$ = .05 and effect size $f$ = .10 were assumed to be the minimum effect size of interest. We recruited 1,427 participants in March 2023 using CloudResearch Connect, an online participant pool that aggregates multiple market research platforms \cite{litman2017turkprime}. 
Participants were all from the United States and were required to be over 18 years old. We also filtered out participants who indicated they were non-partisans. Participants who identified as ``True Independents'' who leaned neither toward the Democratic nor Republican parties were filtered out of the study before participants were randomly assigned to conditions. After ruling out people who failed the embedded attention check question ($n$ = 26), were under the age of 18 ($n$ = 1), used duplicate IP addresses ($n$ = 15), gave incomplete answers ($n$ = 1), or spent less than two minutes on the survey ($n$ = 4), 1,380 participants remained in the data analysis.

\subsubsection{Platform Development}
In order to review the effects of ranking political content in realistic social media environments, we created the PolitiFeed web page that resembles modern social media feeds and used real Facebook posts from political candidates as content for this mock social media feed. This webpage was developed using Flask, a Python web framework, and included study filtering and attention checks as well as the mock social media feed. The site was deployed on Heroku, a cloud-based deployment platform, for easy accessibility and stability when handling a significant influx of subjects. The data was recorded via Qualtrics and included participants' responses and engagement with the political content presented on the web page. 

\begin{figure*}[!t]
	\centering
         \includegraphics[width=\textwidth]{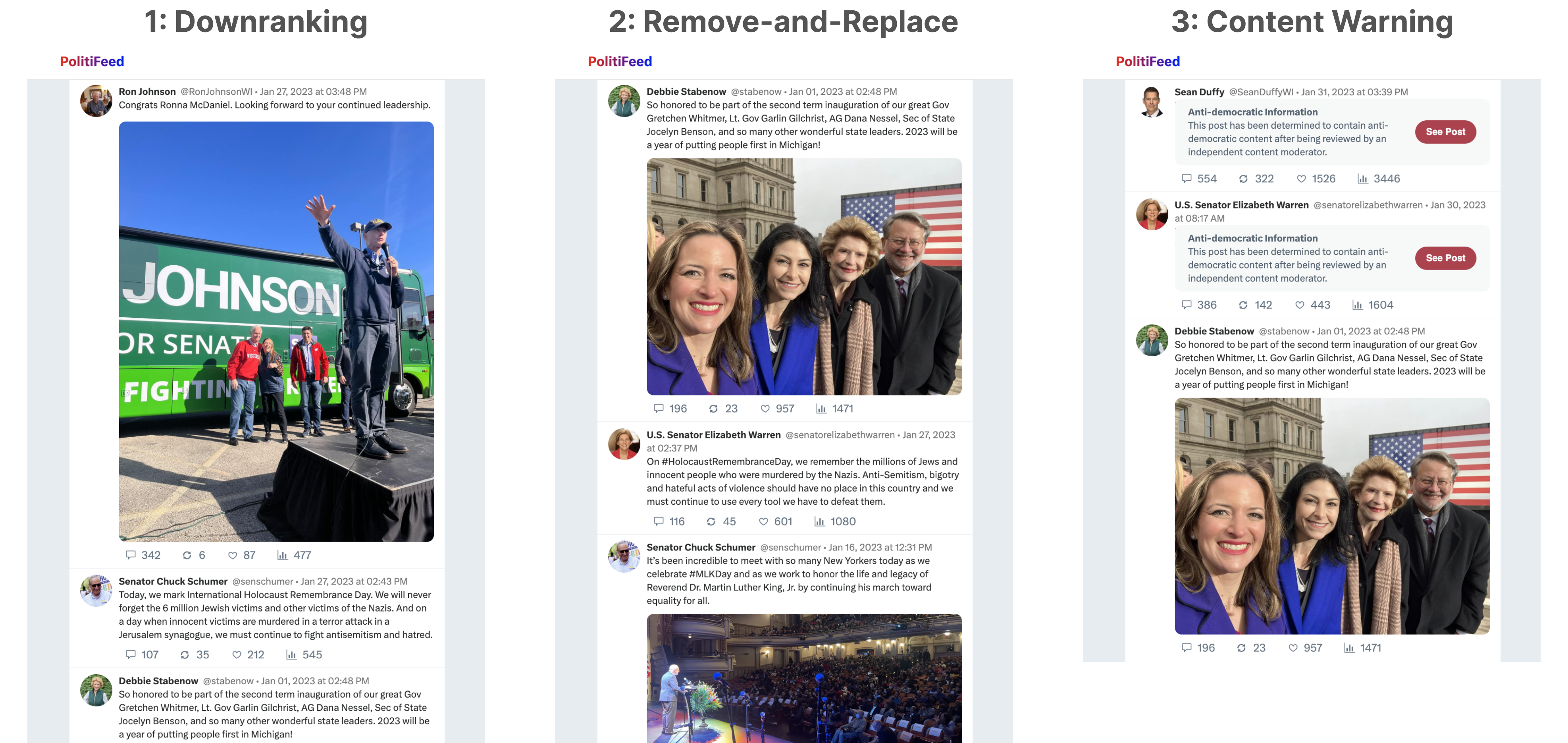}
	\caption{Website Interface of Democratic Attitude Feeds. Participants in different feed conditions were exposed to different interfaces. (Left) Example posts towards the top of the Downranking condition where anti-democratic information is ranked towards the bottom of the feed. (Center) Example posts in the Remove-and-Replace feed in which anti-democratic posts are replaced with pro-democratic posts. (Right) Example posts in the Content Warning feed where anti-democratic posts are blurred with content warnings and users must click the post to see the information.} 
	\label{fig:engagement}
\end{figure*}

We designed PolitiFeed to resemble modern social media platforms to emulate users' behavior and experiences on their own social media feeds and better measure the impacts of alternative ranking methods. Additionally, we intentionally designed the website logo ``PolitiFeed'' with a gradient color (both blue and red) to limit potential political bias. An example of the website interface is shown in Figure \ref{fig:engagement}.

\subsubsection{Measurements}
Here, we provide an overview of our Study 1 measures.

\begin{itemize}
    \item \textbf{Partisan Animosity}. Adapting from \cite{voelkel2023megastudy}, we measured partisan animosity using an instrument called the ``feeling thermometer'' ($M$ = 74.67, $SD$ = 22.26). We asked participants to rate their feelings towards both Democrats and Republicans on a 101-point scale. Participants were told, ``Ratings between 50 degrees and 100 degrees mean that you feel favorable and warm toward them. Ratings between 0 degrees and 50 degrees mean that you don't feel favorable toward them and that you don't care too much for them. You would rate them at the 50 degree mark if you don't feel particularly warm or cold toward them.'' We then reverse coded participants' rating towards outpartisans so that higher scores indicated higher partisan animosity.

    \item \textbf{Support for Undemocratic Practices}. Adapting from \cite{voelkel2023megastudy}, we asked participants to rate their support for undemocratic practices (Cronbach's $\alpha$  = .76, $M$ = 15.62, $SD$ = 18.51) on a 101-point scale by indicating the extent to which they disagree or agree with each of the following statements:
        \begin{itemize}
            \item (Republicans/Democrats) should reduce the number of polling stations in areas that support (Democrats/Republicans).
            \item (Republican/Democratic) governors should ignore unfavorable court rulings by (Democrat/Republican) -appointed judges.
            \item (Republican/Democratic) governors should prosecute journalists who accuse (Republican/Democratic) politicians of misconduct without revealing sources.
            \item (Republicans/Democrats) should not accept the results of elections if they lose.
        \end{itemize}
    
   \item \textbf{Perceived Threat to Freedom of Speech}. Adapting from previous research \cite{dillard2005nature, moyer2010explaining}, four items were included to assess the degree to which one perceives a threat to freedom in response to the ranking methods. Participants were asked to indicate how much they agree or disagree with the following statements based on a 7-point scale (1 = strongly disagree, 7 = strongly agree): ``The social media platform I just used threatened users’ freedom to express on social media,'' ``The social media platform I just used tried to manipulate users on the platform,'' ``The social media platform I just used tried to pressure users on the platform,'' and ``The social media platform I just used tried to make a decision for users.'' The four items were highly correlated and could be formed into a reliable index (Cronbach's $\alpha$ = .93, $M$ = 2.79, $SD$ = 1.57).

    \item \textbf{Feed-level Satisfaction}. We adapted feed-level measures from \citet{jannach2016recommendations} and a survey conducted by Facebook in 2019. After exposure to the holistic social media feed, participants were asked to rate three items: ``Is PolitiFeed you just read worth your time?'', ``Are you satisfied with PolitiFeed?'', and ``Is PolitiFeed you just read helpful for users to find relevant items?'' on 7-point scales from 1 (not at all) to 7 (very). The three items were highly correlated and could be formed into a reliable index (Cronbach's $\alpha$  = .91, $M$ = 4.74, $SD$ = 1.95).
       
   \item \textbf{Time Spent on the Social Media Feed}. We measured engagement level by recording participants' time elapsed on the social media feed ($M$ = 414.51 seconds).
      
    \item \textbf{Post-survey Measurements}. We collected demographic information from participants, such as partisanship strength, education, income, gender, and social media use. 

\end{itemize}

\subsubsection{Data Analysis Plan}
We used multiple generalized linear models (GLMs) with post hoc tests using Bonferroni correction to compare the effects of different ranking methods on participants’ partisan animosity, engagement level, and feed-level satisfaction.

\subsection{Results}

\subsubsection{Democratic attitude feeds significantly reduced partisan animosity}

\begin{figure*}[!t]
	\centering
	\mbox{
		\subfigure[Partisan Animosity (Democrats)
		\label{neu_lib}]{\includegraphics[width=0.49\textwidth]{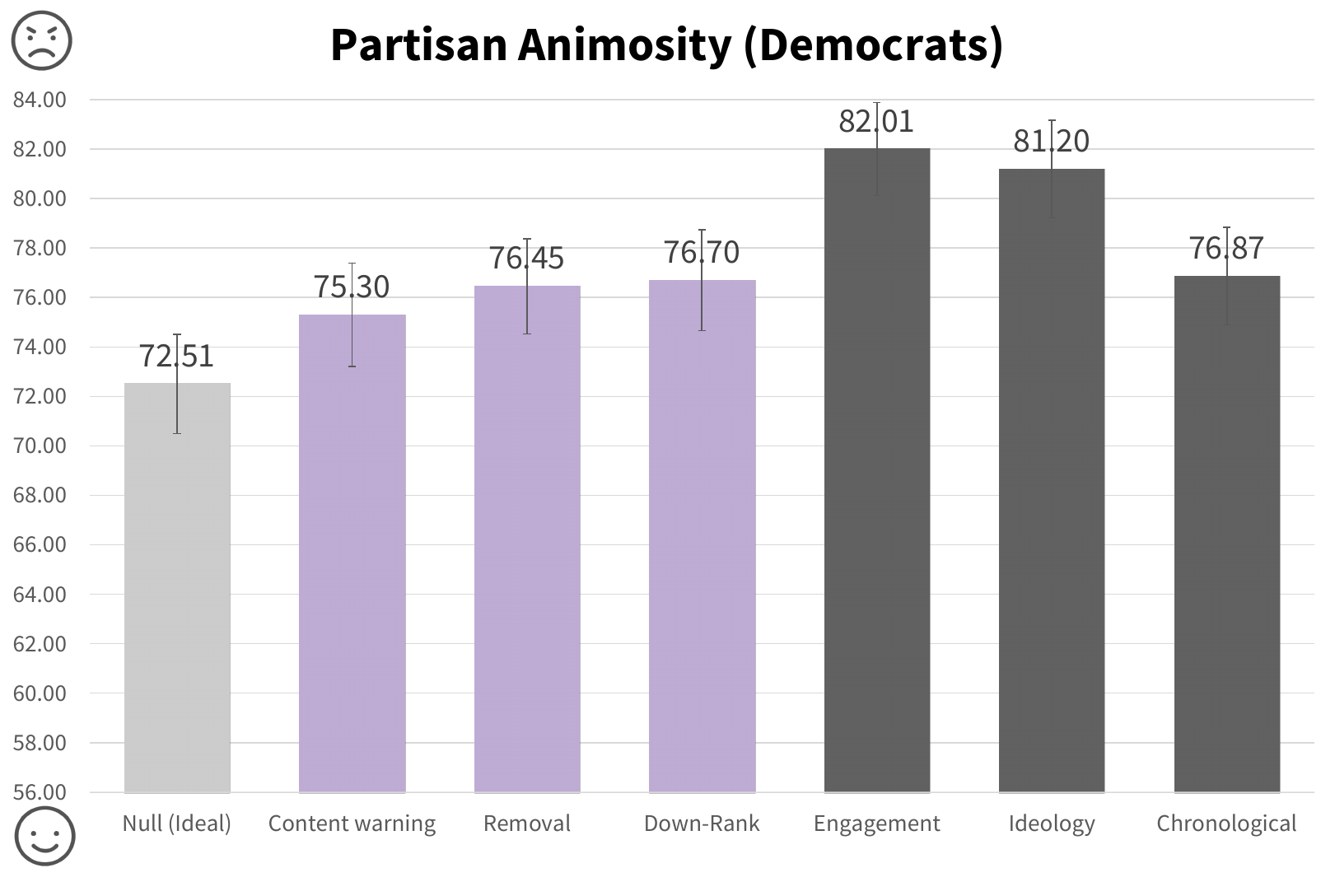}}
		
		\subfigure[Partisan Animosity (Republicans) 
            \label{rev_lib}]{\includegraphics[width=0.49\textwidth]{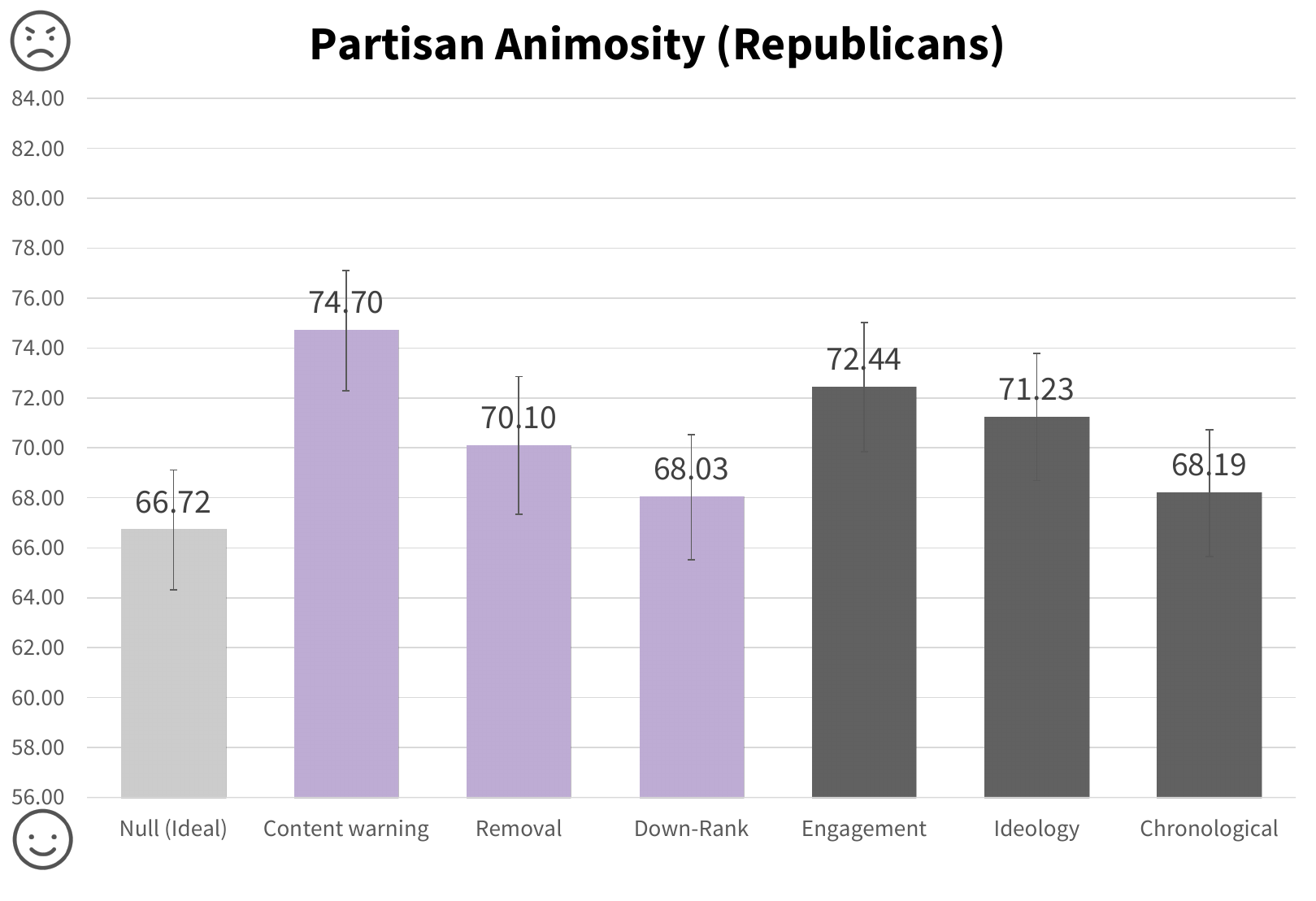}}
		}
	\vspace{-0.15in}
	\caption{Means of Partisan Animosity Across Conditions (Divided by Parties). We found that democratic attitude feeds (in purple)---specifically the downranking feed and remove-and-replace feed --- caused significantly less partisan animosity than the engagement feed (in dark grey).} 
	\label{fig:animosity}
\end{figure*}

\textbf{H1} was mostly supported: it predicted that partisans exposed to the downranking, content warning, and removal conditions will reduce their partisan animosity compared to partisans exposed to the engagement feed and chronological feed. We visualize the results in Figure \ref{fig:animosity}. A generalized linear model (GLM) was used to test the effects of different feeds and party affiliation on the dependent variable, partisan animosity. Both feed condition and party affiliation were entered into the model as independent variables. There was a significant main effect of feed condition, \textit{F}(6, 1366) = 2.68, $p$ = .01, \begin{math}\eta_{\text{p}}^{2}\end{math} = .01, and a significant main effect of party, $F$(1, 1366) = 33.93, $p$ < .001, \begin{math}\eta_{\text{p}}^{2}\end{math} = .02 on partisan animosity. There was no significant interaction effect between condition and party, $F$ (6, 1366) = 1.04, $p$ = .40, \begin{math}\eta_{\text{p}}^{2}\end{math} = .01, which means the effect of the feed condition on partisan animosity did not differ by partisanship. 

We report a full set of pairwise comparisons between conditions in Table \ref{tab:pairwise}.
Notably, the removal and downranking conditions resulted in significantly less partisan animosity than the traditional engagement-based feed ($d = -.20$, $p$ = .04, and $d = -.25$, $p$ = .02). The effect of the content warning condition was only marginal ($d = -.18$, $p = .10$) --- as seen in Figure~\ref{fig:animosity}, it was roughly as effective as the other democratic attitude conditions for Democrats, but far less effective for Republicans. The three democratic attitude feeds had no measurable difference from the chronological condition (all $p >.05$). 
In summary, while the content warning feed backfired with conservatives, the removal and downranking feeds decreased partisan animosity compared to the engagement feed.

We additionally examined the possibility of differential impacts of the feed condition across strong versus weak partisans. We conducted exploratory analyses (that were not pre-registered) by adding partisanship strength as a moderator to the aforementioned generalized linear model. We found a significant moderation effect of partisanship strength on partisan animosity, $p$ = .03. Specifically, the significant difference between the removal and engagement feeds primarily came from weak partisans, $d = -.42$, $p$ = .01, rather than strong partisans, $d = -.14$, $p$ = .51. Similarly, we found the significant difference between the downranking and engagement conditions also came from weak partisans, $d = -.29$, $p$ = .04, instead of strong partisans, $d = -.22$, $p$ = .24.

\begin{table}[tb]
\resizebox{0.65\textwidth}{!}{%
\begin{tabular}{@{}lccccccc@{}}
\toprule
\textbf{Condition Comparisons} & \multicolumn{1}{c}{\textbf{Mean Diff.}} & \multicolumn{1}{c}{\textit{\textbf{SE}}} & \multicolumn{1}{c}{\textit{\textbf{p}}} & \multicolumn{1}{c}{\textbf{Cohen's $d$}} \\ \midrule
Content warning/Engagement                         & -3.21                                        & 1.97                                     & .10                                     & -.18                                            \\
Removal/Engagement                                 & -4.48                                        & 2.12                                     & .04*                                    & -.20                                            \\
Downranking/Engagement                                & -4.97                                        & 2.11                                     & .02*                                   & -.25                                            \\
Null/Engagement                                    & -8.07                                        & 2.10                                     & \textless{}.001***                      & -.40                                            \\
Chronological/Engagement                           & -4.82                                        & 2.10                                     & .02*                                    & -.23                                            \\
Ideology/Engagement                                & -0.95                                        & 2.08                                     & .65                                     & -.06                                            \\
Content warning/Chronological                      & 1.69                                         & 2.17                                     & .44                                     & .07                                             \\
Removal/Chronological                              & .39                                         & 2.32                                     & .87                                     & .03                                             \\
Downranking/Chronological                             & -.17                                        & 2.31                                     & .94                                     & -.02                                            \\
Null/Chronological                                 & -3.23                                        & 2.29                                     & .16                                     & -.15                                            \\
Ideology/Chronological                             & 3.85                                         & 2.28                                     & .09†                                    & .17                                             \\
Content warning/Ideology                           & -2.12                                        & 2.15                                     & .33                                     & -.11                                            \\
Removal/Ideology                                   & -3.49                                        & 2.30                                     & .13                                     & -.13                                            \\
Downranking/Ideology                                  & -4.00                                        & 2.29                                     & .08†                                    & -.18                                            \\
Null/ideology                                      & -7.05                                        & 2.27                                     & .002**                                  & -.32                                            \\
Content warning/Null                               & 4.98                                         & 2.16                                     & .02*                                    & .23                                             \\
Removal/Null                                       & 3.73                                         & 2.31                                     & .11                                     & -.18                                            \\
Downranking/Null                                      & 3.02                                         & 2.30                                     & .19                                     & .14                                             \\
Content warning/Downranking                           & 1.95                                         & 2.18                                     & .37                                     & .08                                             \\
Removal/Downranking                                   & .58                                          & 2.34                                     & .80                                     & .05                                             \\
Content warning/Removal                            & 1.02                                         & 2.20                                     & .64                                     & .03                                             \\ \bottomrule
\end{tabular}
}
\vspace{0.05in}
\caption{Pairwise Comparison of Partisan Animosity (Note: †$p$ < .10, *$p$ < .05, **$p$ < .01 *** $p$ < .001). For each row, the mean difference is calculated by subtracting the partisan animosity of the second condition from that of the first condition.}
\label{tab:pairwise}
\end{table}

\textbf{H2}, which tested the impact on support for undemocratic practices, was not supported; however, this is consistent with prior work, which found that support for undemocratic practices is more resistant to short-term interventions than partisan animosity~\cite{voelkel2023megastudy}. A generalized linear model (GLM) was used to test the effects of different feeds and party affiliation on the dependent variable, support for undemocratic practices. Two independent variables, feed condition and party affiliation, were entered into the model as IVs. There was no significant interaction between feed condition and party, $F$ (6, 1366) = 1.11, $p$ = .36,  partial \begin{math}\eta_{\text{p}}^{2}\end{math} = .01 on support for undemocratic practices, and no main effect of feed condition, $F$ (6, 1366) = .93, $p$ =.47, partial \begin{math}\eta_{\text{p}}^{2}\end{math} = .004. There was a significant main effect of party affiliation, $F$ (1, 1366) = 59.64, $p$ < .001, partial \begin{math}\eta_{\text{p}}^{2}\end{math} = .04 on support for undemocratic practices. Specifically, Republicans ($M$=20.52, $SD$= 21.0) had significantly higher support for undemocratic practices than Democrats ($M$=12.73, $SD$ = 16.10), $p$ < .001, Cohen’s $d$ = .42. Pairwise comparisons showed no significant differences across different feeds. 

\subsubsection{Democratic attitude feeds did not compromise feed-level satisfaction and engagement}
\begin{figure*}[!t]
	\centering
	\mbox{
		\subfigure[Feed-level Satisfaction (Democrats)
		\label{neu_lib}]{\includegraphics[width=0.49\textwidth]{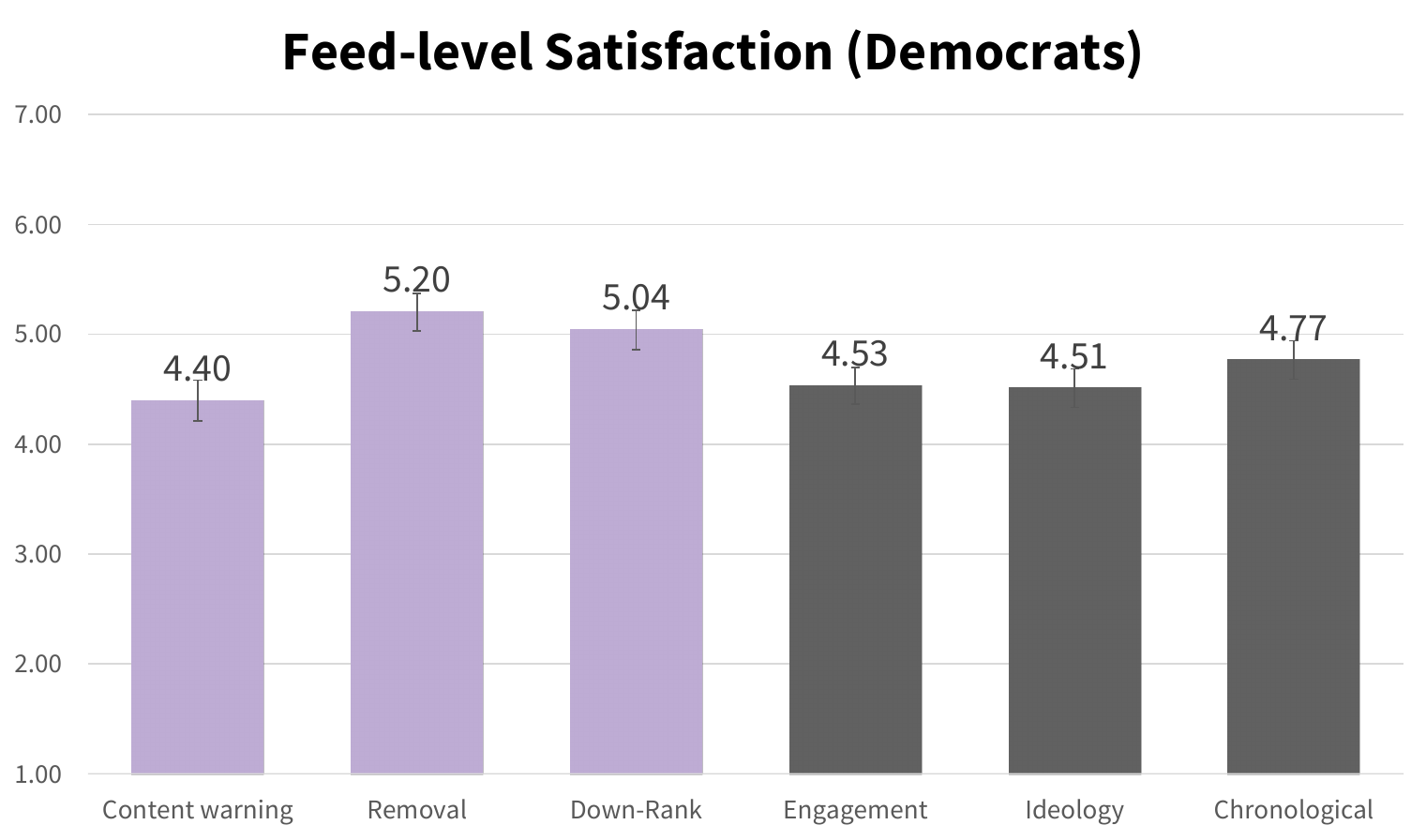}}
		
		\subfigure[Feed-level Satisfaction (Republicans)
            \label{rev_lib}]{\includegraphics[width=0.49\textwidth]{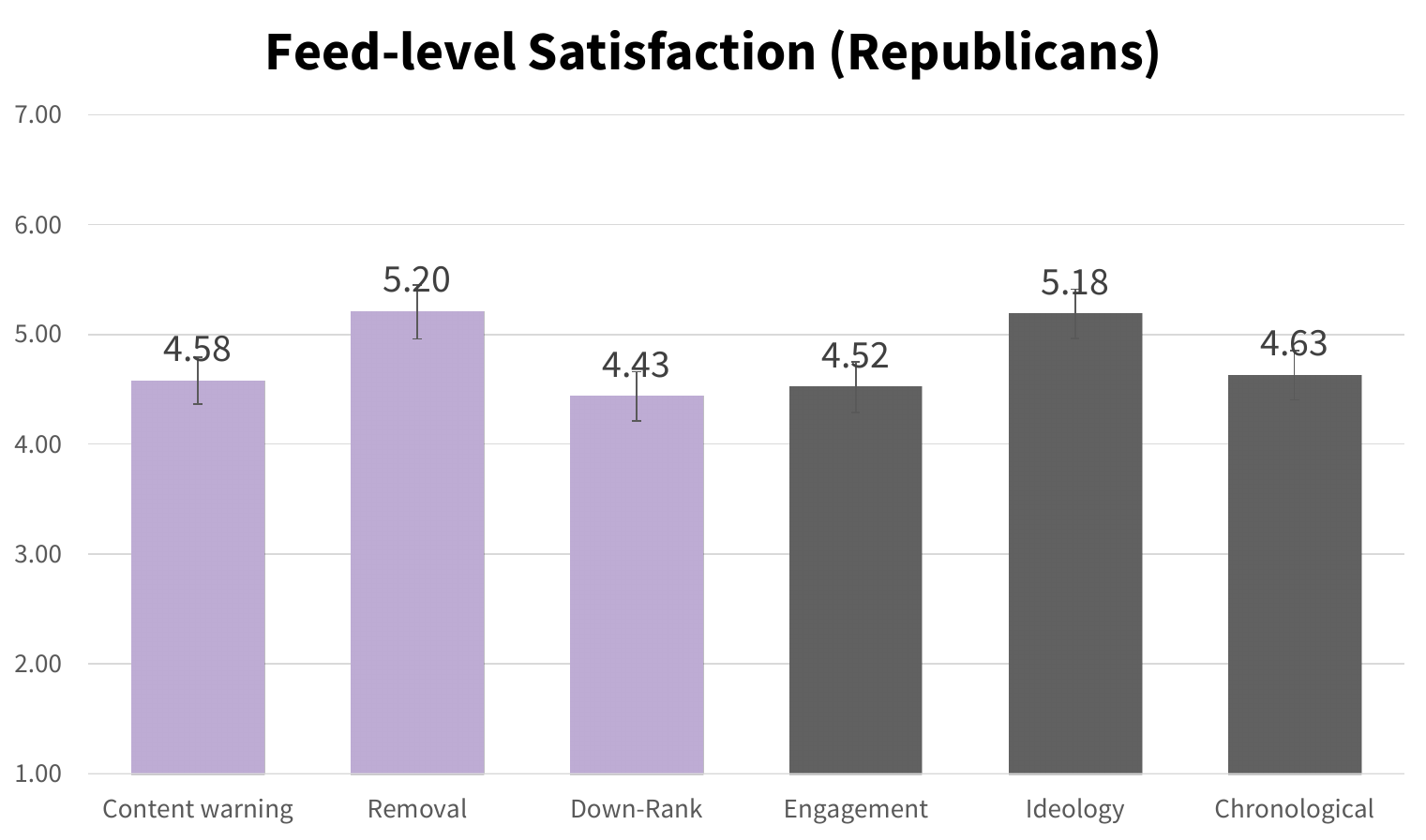}}
		}
	\vspace{-0.15in}
	\caption{Means of Feed-level Satisfaction Across Conditions (Divided by Parties). We found that Democrats exposed to democratic attitude feeds (in purple)---specifically the downranking and removal feed --- had significantly higher feed-level satisfaction than those exposed to the engagement-based feed. Republicans exposed to the removal feed had significantly higher satisfaction than those in the engagement-based feed.} 
	\label{fig:satisfaction}
\end{figure*}

The downranking feed did not compromise participants' satisfaction with the feed. There was a marginal significant interaction between feed condition and party, $F$ (5, 1163) = 2.14, $p$=.06,  partial \begin{math}\eta_{\text{p}}^{2}\end{math} = .01 on feel-level satisfaction. There was a significant main effect of feed condition, $F$ (5, 1163) = 3.15, $p$ = .008, partial \begin{math}\eta_{\text{p}}^{2}\end{math} = .013 on feed-level satisfaction, but no main effect of party affiliation, $F$ (1, 1163) = .02, $p$=.89, partial \begin{math}\eta_{\text{p}}^{2}\end{math} = .000. In fact, Democrats exposed to the downranking and removal-and-replace feed even reported greater satisfaction with the feed compared to those exposed to the feed ranked by engagement (downranking vs. engagement: $p$ = .04, Cohen’s $d$ = .25; removal vs. engagement: $p$ = .005, Cohen’s $d$ = .35). For Republicans, there was no significant difference between their satisfaction with the downranking and the engagement-based feed, $p$ =.79. Republicans in the removal feed showed significantly higher level of satisfaction compared to those in the engagement feed, $p$ = .039, Cohen’s $d$ = .36, as shown in Figure \ref{fig:satisfaction}. 
When comparing to the chronological feed, partisans exposed to the removal feed had significantly higher feed-level satisfaction compared to those exposed to the chronological feed, $p$ = .017, Cohen’s $d$ =.13, but there was no significant difference between the downranking and chronological feed, $p$ = .84.

In addition, we found no significant difference between the time spent on feed when partisans were assigned to a downranking feed ($M$=488.52) and the engagement-based feed ($M$=400.24), $p$ = .30, Cohen’s $d$ = .11 (in seconds). Detailed means and $SD$s are reported in the Appendix ~\ref{tab:mean}.

\subsubsection{Content warning feed prompted freedom of speech concerns}
\textbf{H3}, testing threats to freedom of speech, was supported: the removal and downranking feeds did not prompt freedom of speech threats in participants, but the content warning feed did. A generalized linear model (GLM) was used to test the effects of different feeds and party affiliation on the dependent variable, perceived threat to freedom of speech. Similarly, both feed condition and party affiliation were entered into the models as IVs. There was a marginally significant interaction effect between condition and party, $F$ (5, 1163) = 2.13, $p$ =.06,  partial \begin{math}\eta_{\text{p}}^{2}\end{math} = .01. There was a significant main effect of the feed condition, $F$ (5, 1163) = 16.86, $p$ <.001,  partial \begin{math}\eta_{\text{p}}^{2}\end{math} = .07, and a significant main effect of party, $F$ (5, 1163) = 8.25, $p$ =.004,  partial \begin{math}\eta_{\text{p}}^{2}\end{math} = .01 on perceived threat to freedom of speech. Pairwise comparisons showed that partisans exposed to the content warning feed perceived a significantly higher level of threat to freedom of speech compared to partisans exposed to other feeds, $p$ <.001, with a range of sizable effects (vs. chronological: Cohen’s $d$ = .62; vs. downranking: Cohen’s $d$ =.75; vs. engagement: Cohen’s $d$ =.72; vs. removal-and-replace: Cohen’s $d$ =.77; vs. ideology: Cohen’s $d$ =.57). We suggest that this threat was likely the reason that the content warning condition was less effective than the other democratic attitude feeds in reducing partisan animosity.

\subsubsection{Summary}

Conditions that used manual annotations to craft democratic attitude feeds---the downranking feed and remove-and-replace feed---significantly reduced partisan animosity without reducing engagement and satisfaction, and without raising freedom of speech concerns. Notably, this effect was observed primarily with weak partisans and not with strong partisans, whose strong attitudes are unlikely to be easily modified. Content warnings, using the same manual annotations, did prompt freedom of speech concerns amongst conservatives, and had no overall effect on reducing partisan animosity. None of the approaches impacted support for undemocratic practices.
\section{Study 2: Replicating Manual Annotations with a Large Language Model}

Next, we sought to explore whether we could use algorithmic methods to replicate the effects observed in Study 1. While we found that manual expert annotations served as an effective basis for downranking to mitigate partisan animosity in social media feeds, it is not feasible for manual annotations to serve real-world social media ranking systems. Thus, we turned to large language models (in particular, the GPT-4 model) to investigate how faithfully we could mirror expert annotations with automated approaches that could scale to re-rank social media feeds in production.

\subsection{Democratic Attitude Rating Using GPT-4}

Recent advances in large language models (LLMs) present an opportunity to carry out a broad range of classification tasks purely through natural language instructions. These ``zero-shot'' capabilities---the ability of models to perform tasks without requiring any additional training examples---grant a great deal of flexibility to communicate nuanced constructs~\cite{kojima2022zeroShot, ouyang2022rlhf}. While more traditional deep learning models require substantial technical expertise and effort to gather data and train a performant task-specific model, zero-shot prompting substantially lowers the barrier by only requiring a verbal descriptions of the task and its requirements.

We thus focus on understanding the performance of LLMs for democratic values ranking: if zero-shot modeling with LLMs could sufficiently replicate manual annotations, it would present a path forward to scale up the effect of manual downranking while maintaining accessibility for stakeholders to adjust the values embedded in a social media platform.

\subsubsection{Method}
We leverage GPT-4---at the time of writing, the state-of-the-art large language model developed by OpenAI~\cite{openai2023gpt4}---to explore the feasibility of automating our democratic attitude model. GPT-4 is a transformer-based model that is pretrained to predict the next token in a document and further fine-tuned to generate outputs aligned with human preferences. Owing to the massive scale of data and parameters with which LLMs have been trained, researchers have found that these models display ``in-context learning'' capabilities, whereby a model can carry out a task given only an input of natural language instructions and a few demonstrations of the task~\cite{brown2020fewShot}. Subsequently, instruction tuning methods like those used in GPT-4 have improved LLM capabilities to follow human instructions with methods like reinforcement learning from human feedback (RLHF)~\cite{ouyang2022rlhf}. These advancements have improved LLM performance on zero-shot learning, which omits demonstrations and relies purely on the model's understanding of natural language instructions. Earlier models like GPT-3 and GPT-3.5 have displayed promising performance in applying predefined qualitative codebooks or labeling policies to classify text data. For example, prior work has achieved moderate to substantial agreement with human labels in social science tasks such as political ideology detection, misinformation classification, question analysis, and implicit hate speech detection~\cite{xiao2023qualitative, ziems2023large, huang2023chatgptAnnotation, do2022augmentedSocialScientist}.

To develop our zero-shot prompts, we draw from the definitions provided by~\citet{voelkel2023megastudy} as well as the codebook that our own research team developed in Study 1 to perform manual annotation of the social media post inventory. We split this inventory into a development set (n=205) and a test set (n=200); we used only the development set to iterate on and refine our prompts and reserved the test set for evaluation after we arrived at our finalized set of prompts.

For each anti-democratic variable, we use a prompt of the following format to annotate a social media post:
\begin{lstlisting}[language=Markdown]
    [System Message]
    Please rate the following message's {variable name} from 1 to 3. 
    {variable name} is defined as {variable definition}. 
    Your rating should consider whether the following factors exist in the following message:
        
    {Variable factors: a list of factors relevant to the variable and how they map to the 3-point scale}
        
    After your rating, please provide reasoning in the following format: 
    Rating:__ ### Reason: __ (### is the separator)

    [User Message]
    {social media post content}
\end{lstlisting}
The model then returns an ``Assistant Message'' that we then parse. The variable factors differ for each anti-democratic variable and are listed in full in Appendix~\ref{appendix:prompts}. For a sample of the general format, the variable factors for the ``support for undemocratic practices'' variable are as follows:
\begin{lstlisting}[language=Markdown]
    A: Show support for undemocratic practices
    B1: Partisan name-calling
    B2: Emotion or exaggeration

    Rate 1 if doesn't satisfy any of the factors
    Rate 2 if doesn't satisfy A, but satisfies B1 or B2
    Rate 3 if satisfies A, B1 and B2
\end{lstlisting}

\subsubsection{Implementation details}
We use the OpenAI Python API for GPT-4 to perform our zero-shot modeling task. We use the \texttt{gpt-4-0314} chat completion model, which was the state-of-the-art version at the time of development, and we use a temperature setting of 0.7, consistent with other work that aims to emulate human annotations using LLMs~\cite{pangakis2023automated, hamalainen2023synthetic}.

\subsection{Results}
First, we compared the \textit{overall} ranking results (the total score ranging from 8 to 24) produced by manual ratings versus those produced by automated ratings. We evaluated on our held-out test set ($n = 200$) to understand the holistic impact of the two rating methods on social media feed ranking. 
To capture the overall ranking similarity, we considered two metrics. First, Spearman's rank correlation coefficient, Spearman's $\rho$, is a non-parametric measure of rank correlation that is used to assess the strength of and direction of a monotonic relationship between two variables. Meanwhile, Krippendorff's $\alpha$ is a measure of the agreement between pairs of ratings and is often used in content analysis to assess inter-coder agreement; while other agreement measures such as Cohen's $\kappa$ are designed for categorical variables, Krippendorff's $\alpha$ can be applied to ordinal variables like our anti-democratic attitude scores. We used both metrics to capture the similarity in the final feed ranking order as well as the agreement of the cumulative scores between GPT-4 and manual rating methods.

We observed that the automated ranking results produced by GPT-4 highly correlated with the manual ranking provided by human annotators (Spearman's $\rho$= .75, $p$ <.001, Krippendorff's $\alpha$ = .78). This result suggests that, at the omnibus level of the overall anti-democratic attitude construct, the prompts succeed in closely mirroring the manual labels. These results are consistent with that of prior work using LLMs, which have reported Spearman's $\rho$ values ranging from $.54$~\cite{yang2023large} to $.68$~\cite{kim2023aiaugmented}. We also repeated our procedure with GPT-3.5, the model version that preceded GPT-4, and find comparable performance results, as documented in Appendix~\ref{appendix:gpt-3.5}.

\begin{figure}[!t]
    \centering
    \includegraphics[width=0.5 \textwidth]{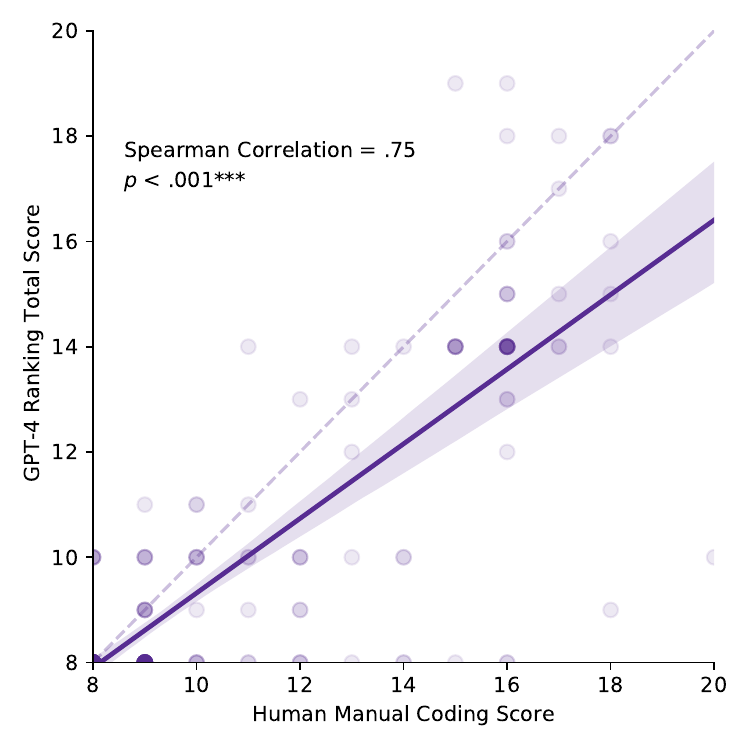}
    \caption{Correlation between manual ranking and GPT-4 ranking. We observe a strong Spearman's rank correlation between the two ranking methods. The GPT-4 rankings skew slightly lower than the manual rankings. The datapoint opacity indicates the number of posts that received the corresponding total-score values.}
    \label{fig:correlation}
\end{figure}

\begin{table}[!t]
\centering
\resizebox{\textwidth}{!}{
\begin{tabular}{lccc}
\toprule
\textbf{Individual anti-democratic attitude variables}              & \textbf{Krippendorff's $\alpha$} & \textbf{Classification Accuracy} & \textbf{F1 Score} \\ 
\midrule
Partisan Animosity                     & .806                    & .905                    & .711                        \\
Support for Undemocratic Practices     & -.093                   & .725                    & .303                        \\
Support for Partisan Violence          & .730                    & .900                    & .865                        \\
Support for Undemocratic Candidates    & .321                    & .975                    & .496                       \\
Opposition to Bipartisanship           & .689                    & .890                    & .561                        \\
Social Distrust                        & .608                    & .790                    & .611                        \\
Social Distance                        & .685                    & .775                    & .551                       \\
Biased Evaluation of Politicized Facts & .507                    & .560                    & .515                       \\ 
\bottomrule
\\
\toprule
\rowcolor{purple}
\textbf{Outcome variable}              & \textbf{Krippendorff's $\alpha$} & \textbf{Spearman's $\rho$} & \textbf{Mean Absolute Error (MAE)} \\ 
\midrule
\rowcolor{purple}
\textbf{Overall democratic attitude ranking (8-24 scale)}                     & .778                    & .754     & 1.235               \\
\bottomrule
\end{tabular}
}
\vspace{0.05in}
\caption{Performance metrics for GPT-4 ratings on the overall democratic attitude ranking and individual anti-democratic attitude variables. We observe a strong correlation between the overall manual and GPT-4 feed ranking results.
}
\label{tab:gpt_allVars}
\end{table}




Then, diving into the \textit{individual} anti-democratic attitude variables, we observed substantial agreement between GPT-4 ratings and manual ratings, with a median Krippendorff's $\alpha$ of $.647$, as shown in Table~\ref{tab:gpt_allVars}. 
Treating the 3-point scale options as a multiclass classification task, we observed a mean classification accuracy of $.815$. 
Since the manual ratings for all variables were skewed towards the negative class (across variables, a mean of 79.2\% examples received a manual rating of 1), we also calculated the F1 score, which is a more reliable indicator of model performance for imbalanced datasets. We observed a mean F1 score of $.577$ across variables. 
The only two individual variables that displayed Krippendorff's $\alpha$ and F1 scores below $.5$ were \textit{support for undemocratic practices} and \textit{support for undemocratic candidates}. Based on manual ratings, both of these variables displayed extremely large skew towards the negative class such that 99\% of examples in the test set were manually annotated with a score of 1. This skew made it challenging to calibrate the model and resulted in lower performance results. However, the large skew on those variables meant that even in the manually-ranked feeds, these two variables most often had a value of 1 and contributed less to the overall democratic attitude score than the other variables that displayed greater variation. Thus, misalignments in the GPT-4 ratings for the highly skewed variables did not appear to substantially affect alignment with manual ratings for the overall feed ranking.
Thus, we concluded that our automated democratic values ranking method using GPT-4 appears to effectively replicate our manual ranking outcomes.

\subsubsection{Summary}
Adapting qualitative codebooks as zero-shot prompts, algorithmic feed ranking using GPT-4 achieved a strong correlation with the manual democratic attitude feed ranking. For individual anti-democratic attitudes, GPT-4 ratings generally aligned well with manual ratings except for two variables, \textit{support for undemocratic practices} and \textit{support for undemocratic candidates}, which were especially rare in the dataset.
\section{Study 3: Replication Using The Democratic Attitude Model}
Closing the loop, we wished to test whether a feed using the automated democratic attitude model labels replicated the effect on partisan animosity that we observed in Study 1 with the manual labels. In Study 3, we conducted a pre-registered replication study ($N$ = 558)\footnote{A priori power analysis using G*Power determined that a total sample of at least 432 participants would be needed to achieve 80\% power if $\alpha$ = .05 and effect size $f$ = .10 were assumed to be the minimum effect size of interest.} among US partisans to compare the algorithmic downranking feed to a manual downranking feed and to a control feed ranked by engagement. Participants were recruited from CloudResearch Connect. Those who participated in Study 1 were ruled out from the study pool. The experimental procedure and measurements remained the same as Study 1.

\subsection{Experimental Design and Hypothesis}
In Study 3, we conducted a between-subjects study design online experiment with three conditions in June 2023. The \textit{manual} condition was the downranking feed from Study 1 that utilized manual expert labels. The \textit{algorithmic} condition was a downranking feed using the output from the democratic attitude model using GPT-4 instead of manual labels. The \textit{engagement} control condition was identical to Study 1. We predicted that partisans exposed to the algorithmic feed would reduce partisan animosity compared to partisans in the engagement feed\footnote{We only wanted to replicate the effect on partisan animosity as we did not find a significant effect on support for undemocratic practices in our Study 1 (H2).}:

\begin{enumerate}
    \item[\textbf{H4:}] Partisans exposed to the (a) manual downranking and (b) algorithmic downranking feed on social media will reduce partisan animosity compared to partisans in the engagement feed.
\end{enumerate}

\subsection{Results}

\subsubsection{Both algorithmic and manual downranked feeds significantly reduced partisan animosity.}
\begin{figure*}[!t]
	\centering
	\mbox{
		\subfigure[Partisan Animosity (Democrats)
		\label{neu_lib}]{\includegraphics[width=0.49\textwidth]{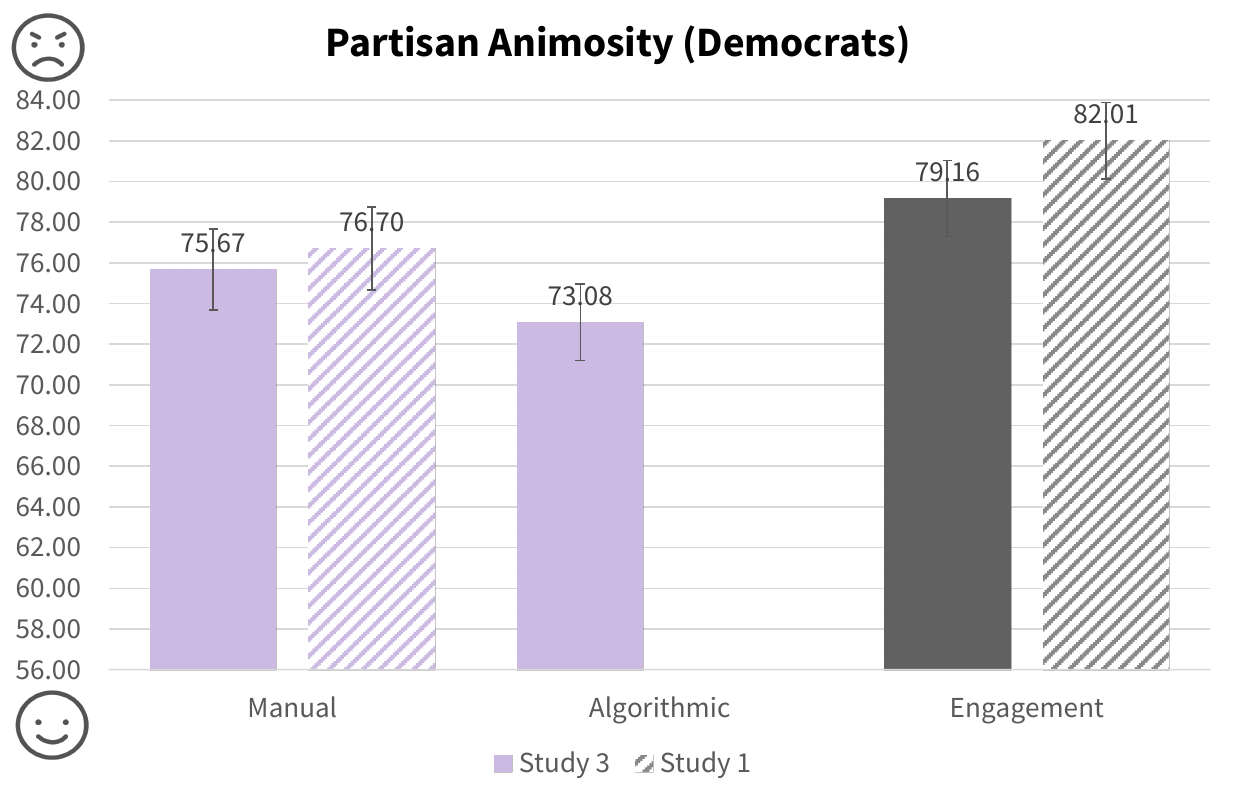}}
		
		\subfigure[Partisan Animosity (Republicans) 
            \label{rev_lib}]{\includegraphics[width=0.49\textwidth]{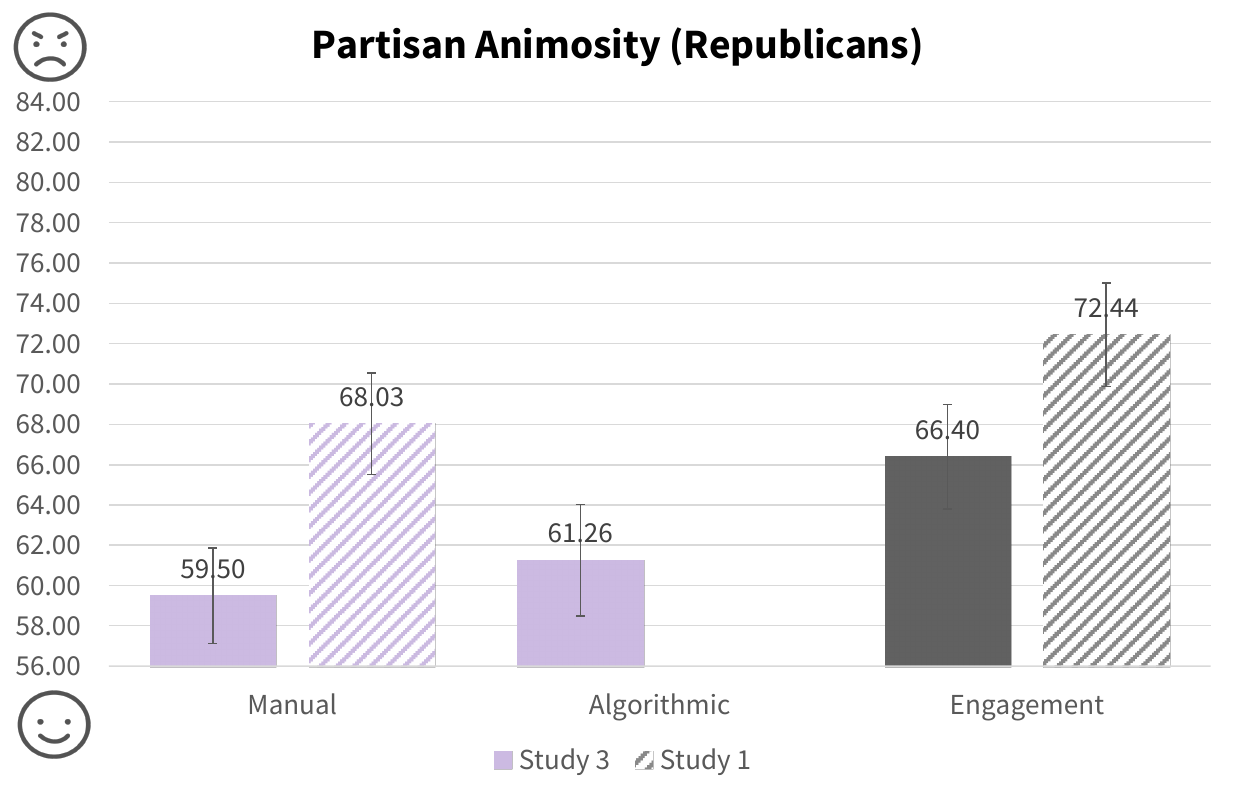}}
		}
	\vspace{-0.15in}
	\caption{Means of Partisan Animosity in Both Study 1 and Study 3 Across Conditions (Divided by Parties). We found that both manual and GPT-4-based downranking feeds caused significantly lower partisan animosity compared to the engagement-based feed. We also found similar effect sizes in Study 1 and Study 3 between the two downranking feeds and the engagement feed (manual vs. engagement: Study 1, $d = -.25$, and Study 3, $d = -.27$; algorithmic vs. engagement: Study 3, $d = -.25$).} 
	\label{fig:animositygpt}
\end{figure*}

A generalized linear model (GLM) was used to test the effects of different feeds and party affiliation on the dependent variable, partisan animosity. Both feed condition and party affiliation were entered into the model as two independent variables. \textbf{H4} was supported (Figure~\ref{fig:animositygpt}): the algorithmic, GPT-4 based democratic attitude feed reduced partisan animosity without impacting time on the site. There was a significant main effect of condition, $F$(2,557) = 3.77, $p$ = .024, and a significant main effect of party, $F$(1,557) = 53.31, $p$ <.001, on  partisan animosity. There was no significant interaction effect between party and condition, $F$(2,557) = .52, $p$ = .60. 

Multiple pairwise comparisons using Bonferroni correction showed that partisans exposed to both the manual ($M$ = 68.96, $SD$ = 23.33) and the algorithmic ($M$ = 69.32, $SD$ = 22.59) downranked feeds displayed significantly lower partisan animosity compared to those exposed to the engagement feed ($M$ = 74.82, $SD$ = 19.73), manual vs. engagement: $p$ = .02, $d$ = -.27, algorithmic vs. engagement: $p$ = .036, $d$ = -.25. Same as in Study 1, there was no significant difference of time spent on the feed across conditions, $F$(1,557) = .26, $p$ = .77, \begin{math}\eta_{\text{p}}^{2}\end{math} = .001 (manual: $M$ = 310.16, $SD$ = 256.96; algorithmic: $M$ = 326.16, $SD$ = 255.01; engagement-based: $M$ = 318.47, $SD$ = 275.05), which indicates that both democratic attitude feeds did not compromise time spent on the feed.

\subsubsection{Additional analyses on strong vs. weak partisans}
As in Study 1, we conducted additional exploratory analyses (not pre-registered) to examine different impacts of feed conditions on strong vs. weak partisans. We again found a significant difference between the downranking and engagement conditions for weak partisans (manual vs. engagement: $d = -.44$, $p$ = .05; algorithmic vs. engagement: $d = -.26$, $p$ = .08), but not for strong partisans (manual vs. engagement: $d = -.18$, $p$ = .81; algorithmic vs. engagement: $d = -.14$, $p$ = .45).

\subsection{Summary}
In summary, Study 3 replicated the original result using algorithmic labeling rather than manual annotation. The effect size between the downranking feeds and the engagement feed on reducing partisan animosity (manual vs. engagement: $d$ = -.27; algorithmic vs. engagement: $d$ = -.25), is also the same as Study 1. Additional analyses suggested that the significant differences between the downranking and engagement conditions come from weak partisans and not strong partisans.
\section{Discussion}
Our work seeks to directly embed democratic values into social media feed algorithms. We focused on political discourse on social media, and our results showed that encoding democratic values into social media feeds could reduce partisan animosity without compromising engagement and satisfaction with the feed. We identified a social science construct and its associated measures, used LLMs to replicate human annotations on those constructs, and redesigned social media feed ranking objectives to account for these LLM annotations. Our results shed light on the impact of feeds that encode societal values, which provides important theoretical and design implications for the field.

\subsection{Theoretical Implications}
The first theoretical contribution of our work is the societal objective function, which seeks to translate social science theory into algorithmic objectives that drive user-facing systems. This sociotechnical approach allows us to embed societal values in social media AIs by translating social science measures of anti-democratic attitudes directly into objective functions. Our work is one of the first studies that not only 1)~identifies the success of translating measures from social science theory into prompts for LLMs such as GPT-4, but also 2)~proves that introducing societal objective functions \textit{into user-facing systems} can result in significant attitudinal changes (i.e., reducing partisan animosity) by conducting empirical experimental studies with users. We envision that this approach can apply to many other domains to translate measures developed by the social scientists in areas such as mental health, well-being, and social equity into different algorithmic objectives. This approach opens up a new area for social media AIs to test and understand how a feed embedding various values might affect users' attitudes and behaviors, along with other outcome variables that social media platforms may care about (e.g., engagement, user experience).

Second, we took a \textit{top-down}, value-centered approach to \textit{intervene} on reducing partisan animosity, which broadens the scope of polarization literature. Prior work tends to take two forms: one body of literature investigates whether today's social media algorithms amplify partisan animosity~\cite{milli2023twitter, tornberg2022digital} or harm democracy~\cite{lorenz2023systematic}. Another line of work investigates the efficacy of bottom-up, ``light-touch'' interventions via websites or social media platforms, such as correcting out-partisan misperceptions through a chatbot~\cite{hartman2022interventions} or deploying interventions like readings, videos, or activities~\cite{voelkel2023megastudy}. However, prior work has not taken a top-down approach to investigate the impact of embedding high-level democratic values into social media AIs. 

Our third theoretical contribution is that our democratic attitude feed could significantly reduce partisan animosity among US partisans without compromising their feed engagement level. This outcome is, of course, dependent on further field and longitudinal studies. However, such results are consistent with past work that indicates that time spent on feed was not in conflict with reducing polarization \cite{saveski2022perspective}. Prior studies have focused on comparisons between feeds ranked in chronological order and feeds ranked by engagement metrics. For instance, one of the recent studies on Facebook during the 2020 US presidential election has examined the effect of substituting the algorithmic feed with a reverse-chronologically-sorted feed \cite{guess2023social}. Extending prior work, our experiments examined seven different feeds that moved beyond chronological and engagement-based ranking to consider alternative feed designs oriented around democratic attitudes, such as a content warning feed that adds friction to viewing anti-democratic information. Furthermore, we see evidence that democratic attitude feeds are more effective in reducing partisan animosity among weak partisans than strong partisans, which suggests that democratic attitude feeds are more likely to depolarize people who are less extreme on the ideological spectrum.
 
In addition to the engagement metric, our work also examines people's perceived threat to freedom of speech and found that partisans exposed to the content warning feed perceived a significant higher level of threat to freedom of speech compared to partisans exposed to other feeds, which is consistent with prior work \cite{dillard2005nature, moyer2010explaining}. We found a backfire effect in the content warning feed: conservatives are more likely to increase partisan animosity in the content warning feed compared to the engagement feed, which suggests that while the content warning condition may nudge people on the decision made by the platform, it may also accidentally create a backfire effect that over-corrects people’s estimation of the percentage of anti-democratic information online.

While this work was under review, a series of recent studies authored by political science researchers in collaboration with Meta were released in Nature and Science. Those studies have also examined the effects of reducing exposure to content from like-minded sources~\cite{nyhan2023like, guess2023social}, reshared content~\cite{guess2023reshares}, or substituting the algorithmic feed with a reverse-chronologically-sorted feed~\cite{guess2023social} during the 2020 US presidential election. For instance, \citet{guess2023social} found little evidence of Facebook’s feed algorithms on altering the levels of affective polarization. How do we draw the line between our study, which found impacts, and theirs, which did not? 

One possibility may be that, while it is challenging to shift long-held attitudes like affective polarization as \citet{guess2023social} found, it may be possible to shift more \textit{immediate} reactions and attitudes, especially when interventions are targeted carefully and in stronger doses. While the feeling thermometer measurement is traditionally treated as a long-term construct~\cite{iyengar2019origins}, our work applies this measurement in a short-term context and finds that people’s feelings towards the opposite party may be movable when we consider the immediate effects of a social media intervention. Additionally, our study enacts a particularly strong intervention: while the Facebook feed typically contains a small percentage of political content~\cite{guess2019accurate}, our feed intervention consists \textit{entirely} of political social media posts. We also deploy this strong intervention on a more targeted sample of partisans rather than general Facebook users. In fact, we find that the participants most influenced by our intervention are weak partisans rather than strong partisans, which supports the notion that stronger, entrenched views may be difficult to shift.

Taken together, these differences in our study design and findings suggest that short-term shifts may be a promising route to enact change. Even if we do not change people’s longitudinal attitudes, short-term interventions have strong potential to reshape how users interact on social media and may allow them to engage in civil dialogue even if they hold differing long standing views.

\subsection{Design Implications}
In ``The Two Cultures'' \cite{snow1959two}, C\@.P\@. Snow famously worried that the sciences were developing a culture wholly separate from the human-centered endeavors. Toward bridging this divide, our work suggests that it is possible to transfer some social science constructs and replicate their effects using a technical system based solely on natural language instruction inputs. These findings carry implications for future technical approaches that might embed societal values in social media and sociotechnical systems more broadly.

\subsubsection{Incorporating societal values in social media}
In this work, we focus on encoding democratic attitudes in social media feeds because these values carry important ramifications for a healthy democracy. Today's social media \textit{already} embeds values. Though implicit, social media AIs hold a de facto position on how democratic attitudes are rendered in the feed, and they similarly hold a de facto position for any other societal value we may choose to study~\cite{bernstein2023embeddingJOTS}. Thus, we believe our societal objective functions approach should be used to experiment with a wide range of other societal values such as mental health, self-expression, diversity, and environmental sustainability~\cite{stray2022building}. Work in this vein would allow us to make important values both explicit and tuneable.

Furthermore, once a broader range of values have been explored and tested with users, we might start to explore feeds that combine different sets of values and better understand how the values trade off against each other and against status quo metrics like engagement and revenue that tend to be more aligned with corporate interests. While we found that encoding democratic attitudes did not have a negative impact on engagement metrics, as we expand the types of values and the combinations of values that we attempt to encode, we likely will encounter tougher tradeoffs between societal values and financial feasibility for platforms. Field experiments will be critical to better understand and navigate these tradeoffs. 

There are also opportunities to intervene in different parts of the social media experience. While we intervened on feed ranking algorithms, societal objective functions might be valuable within post authoring interactions to encourage posters to self-reflect on the values expressed in their posts, or in notification systems to selectively notify users about content that upholds important values.

Finally, it would be valuable to explore the impact of feeds that encode societal values both over longer stretches of time with longitudinal studies and across a diverse set of communities and platforms with expanded field deployments. A benefit of our LLM-based method is that it can scale up to large amounts of content and users, and it can be flexibly adapted to perform ranking for a range of potential platforms.
 
Building on prior work on Value-Sensitive Algorithm Design~\cite{zhu2018valuesensitivealgorithm} which outlined methods for incorporating stakeholder values in a bottom-up manner, our work may also apply to non-social media contexts. Any task that requires content to be classified or scored, for example news recommendation, ad targeting, or toxicity detection, may benefit from integrating societal values. Thus, our approach could potentially help to streamline the workflow of transferring social science findings from more focused, in-lab studies to large-scale deployments in real-world systems. Additionally, since our approach does not require substantial technical expertise to implement and train a custom model, it may help to expand the set of researchers who can test alternative algorithms and interfaces. 

\subsubsection{Technical method: extensions and limitations}
The algorithmic approach taken in our work leverages a large language model to directly generate ratings. Past work has explored the use of LLMs to perform deductive coding that applies existing qualitative codebooks or labeling policies~\cite{xiao2023qualitative, ziems2023large, huang2023chatgptAnnotation, wang2021reducelabeling}. We extend this work to design a codebook for an LLM to provide ratings for a complex concept of anti-democratic attitudes. Furthermore, a key contribution of our work is our societal objective functions method, which not only evaluates the LLM ratings themselves (with comparisons to human ratings), but also investigates the \textit{impact} of the LLM ratings when used to power a user-facing system in an online experiment.
However, future approaches could utilize LLMs in a variety of other ways to achieve more efficient or fine-tuned results. For example, the LLM output could be used to perform knowledge distillation to train a smaller model that may be far less costly or time-intensive to use for inference to serve production systems~\cite{hsieh2023distilling, tang2019distilling}. Alternatively, the LLM-based method could be used to first validate the social science constructs, and experimenters could proceed to build custom models for each construct using more traditional data-driven approaches. 

While the LLM-based method substantially lowers the barrier to entry, there are risks of using such a model for production social computing systems. First, LLM ratings may not always align with desired manual annotations and human judgment, especially if the model is applied to subjective concepts where the human population may disagree~\cite{santurkar2023opinions}. Thus, it is important that model-generated ratings are closely monitored for any real-world deployments or field studies. Large language models may not achieve as high performance and annotator alignment for tasks that require domain-specific knowledge that falls outside of the training data of internet text, and their alignment with current societal values may shift over time. Furthermore, these models encode known biases~\cite{li2021gender, nadeem2021stereoset, sheng2019nlg_bias}, and they may have a variety of biases that are not yet understood that may impact performance. For example, if an LLM displays poor understanding of the sentiment of language used by a particular community, it may struggle to recognize anti-democratic speech from that community. Another risk of directly applying LLM-based ranking models in production systems is that such models could be attacked with adversarial inputs that take advantage of known limitations such as prompt injection~\cite{ignore_previous_prompt} and vulnerabilities similar to those of traditional computer programs~\cite{kang2023exploiting}, so system developers must include safeguards against such practices. 

\subsubsection{Implications for industry practitioners}

Why should social media companies redesign their algorithms? If status quo ranking algorithms are serving these companies well, there may be little incentive to alter their algorithms to incorporate societal values~\cite{ciampaglia2018algorithmic, milli2021optimizing}. While engagement-oriented objectives speak to shorter-term user satisfaction and retention, societal objective functions speak to longer-term impacts on users and society at large. Thus, our method may enable platforms to better evaluate how the design decisions of today impact the viability of their platform far into the future. Furthermore, our work presents initial evidence that societal objective functions may \textit{not} require compromises on the engagement metrics that social media platforms already care about.

If industry practitioners take on our approach, we recommend that they test the impact of manually-annotated social media feeds on real users before implementing any large-scale, algorithmic approaches, as we demonstrated in our work. This gradual, phased approach can help to reduce risks of harm to a platform and its users. Since our study was conducted on a much smaller group of participants than would be exposed on a real social media platform, unforeseen effects may result from large-scale deployments without such precautions.

Finally, given that societal objective functions carry tremendous ethical implications on users and society, we recommend that industry practitioners continuously seek feedback from ethicists, researchers, policymakers, and community stakeholders to better understand the risks and benefits they pose. 

\subsection{Limitations}
We note several limitations of our work that may serve as important directions for future work. First, the participant pool used for our online experiments is limited to partisans based in the United States, so results may vary for U.S. non-partisans and participants based in other countries. In addition, these participants were sourced from a crowdsourcing platform (CloudResearch Connect), so they may not comprise a nationally representative sample of the pool of U.S. partisans, as workers CloudResearch may have relatively higher digital literacy than the general population \cite{yaqub2020effects, jia2022understanding}.

Next, our data source was restricted to only political content sourced from CrowdTangle, which draws from public posts on Facebook. Since political content comprises a relatively small portion of social media feeds (\citet{scharkow2017measurement} and \citet{guess2019accurate} suggest 6\% political tweets and 2\% political Facebook posts), our democratic attitude model may have a subtler effect on real-world feeds that hold a proportionally smaller inventory of political content. Results may also differ for feed posts sourced from different platforms---such as Twitter, Reddit, or Instagram---or for posts authored for a private rather than public audience. In addition, our Study 3 results indicate an overall decrease in partisan animosity across parties and conditions when we replicated Study 1 three months after it had been conducted. We suggest that future work use timely social media posts as stimuli for studies on social media AIs, which are generally time-sensitive.

While we observed that downranking and remove-and-replace democratic attitude feeds reduced partisan animosity, we note that these reductions are comparable to the reductions achieved by a non-algorithmically ranked, reverse-chronological feed. These results suggest that status quo engagement-based ranking may pose the greatest risks to democratic attitudes and that platforms may consider chronological ordering if they turn to non-algorithmic ranking. However, given that platforms may prefer curated ranking approaches in contrast to idiosyncratic chronological ordering~\cite{twitter_reverse_chron}, democratic attitude feeds present a means to mitigate partisan animosity while maintaining curatorial control.

As noted earlier, while our online experiments demonstrated promising results by which democratic attitude feeds could reduce partisan animosity, further experiments are necessary to determine whether such feed interventions could bring about longitudinal effects. We recommend that future work carry out large-scale, longitudinal field experiments that intervene on users' own social media feeds to understand the impact of democratic attitude feeds in the real world.

Finally, our work focused on the societal value of mitigating partisan animosity, but we have not explored other potential societal values. There may be differing challenges in translating other social scientific constructs into societal objective functions, which may require different technical methods and may result in different effects on users in online experiments. An exciting direction of future work will be to explore a range of other societal values to investigate whether and how they can be translated into societal objective functions.

\subsection{Ethics and Societal Impact}
Given the substantial influence that social media platforms hold on the diffusion of information, societal attitudes, and the functioning of democratic processes, we must consider the ethical implications of our work.
First, any form of algorithmic feed ranking inherently poses risks to freedom of expression.
In a democracy, all citizens are allowed the right to express their perspectives and opinions equally \cite{bernholz2021democratictheory}. A social media feed that algorithmically curates social media posts may not capture the full spectrum of perspectives on the platform and, by design, must limit the visibility of certain posts. While our work aims to explicitly foreground societal values that might enhance the diversity of information on social media feeds, our ranking method shares limitations with existing algorithmic ranking methods in that no social media platform can show all posts \cite{sep-freedom-speech}. 

Another risk of our approach is that it may lack nuance on how to handle posts that may appear misaligned with societal values. For example, while partisan animosity in large doses may exacerbate political divides, anger and other negative emotions may be appropriate in the face of injustice; protest and debate is a critical part of political discourse. To avoid this failure mode, societal objective functions must take care to achieve balance among multiple values rather than solely focusing on any one individual value.

Next, our work provides a method to encode societal values in algorithmic objectives, but our method alone does not determine \textit{what values} are encoded or \textit{who decides} what values to encode. Thus, there are risks that the same methods that allow us to encode pro-social societal values could be exploited by bad actors to proliferate harmful content or manipulate ranking algorithms~\cite{torres2022manipulating}. 
We note that today's social media platforms already implicitly or explicitly encode certain societal values that reflect the values of their developers~\cite{seaver2017algorithms}, even if they were not intentionally specified. By making societal values an explicit, measurable goal, our approach provides to good-faith actors both a more realistic understanding of the values encoded in their status quo systems and a means to monitor improvements or adversarial attacks to those values. While we cannot prevent malicious actors from attempting to generate societally harmful content with similar methods, approaches like ours can help platforms to enhance their ``defense'' in detecting and downranking harmful content. Another potential concern is that societal objective functions may have uneven impacts across diverse communities, which may complicate decisions about what values to implement.
Extensive research has demonstrated that social media platforms often contribute to and amplify social inequities~\cite{robinson2015digitalinequities, hargittai2013oxford, lutz2022inequalites}, and even attempts to promote algorithmic fairness may not in fact address the needs of marginalized communities~\cite{gorwa2020algomoderation, hoffmann2019fairness, binns2017fairness}. Continued research on societal objective functions should incorporate the input of marginalized communities and carefully investigate the impacts on these communities to avoid perpetuating harmful disparities.

\section{Conclusion}

Our work embeds the social scientific construct of anti-democratic attitudes into a social media AI objective function. We demonstrate that the survey instruments from prior work on this construct can be adapted into prompts for a large language model (LLM), producing a \textit{democratic attitude model}. Through a series of three studies, we found that social media feeds that integrate this democratic attitude model can significantly reduce partisan animosity without compromising user engagement levels. This \textit{societal objective function} method presents a novel strategy for translating social science theory to algorithmic objectives, which opens up new possibilities to encode societal values in social media AIs.

\begin{acks}
Our research was supported by the Hoffman-Yee Research Grants at Stanford Institute for Human-Centered Artificial Intelligence (HAI). Chenyan Jia was supported in part by Stanford Cyber Policy Center and Stanford Center on Philanthropy and Civil Society. Michelle S. Lam was supported by Stanford HAI and a Stanford Interdisciplinary Graduate Fellowship. We thank other members of our HAI research project, namely Angèle Christin, Jeanne Tsai, Johan Ugander, Nathaniel Persily, Tatsunori Hashimoto, Martin Saveski, Tiziano Piccardi, Chunchen Xu, and Marijn Nura Mado. We also thank our participants who provided valuable insights.
\end{acks}

\bibliographystyle{ACM-Reference-Format}
\bibliography{main_bib}

\section*{Appendix}

\subsection{Inter-rater Reliability}
Table~\ref{tab:manual_irr} displays the inter-rater reliability results between two expert annotators for the eight anti-democratic variables.
\begin{table}[!t]
\centering
\footnotesize
\begin{tabular}{llccc}
\toprule
\textbf{Outcome variable} & \textbf{Krippendorff's $\alpha$} \\ 
\midrule
Partisan Animosity                     & 0.889\\
Support for Undemocratic Practices     & 0.798\\
Support for Partisan Violence          & 0.697\\
Support for Undemocratic Candidates    & 0.798\\
Opposition to Bipartisanship           & 0.810\\
Social Distrust                        & 0.775\\
Social Distance                        & 0.807\\
Biased Evaluation of Politicized Facts & 0.756\\ 
\midrule
Total Score (sum) & 0.895\\
\bottomrule
\end{tabular}%
\vspace{0.05in}
\caption{Inter-coder reliability on the eight anti-democratic variables.}
\label{tab:manual_irr}
\end{table}

\subsection{Additional Tables in Study 1}
Table ~\ref{tab:mean} lists means and $SD$s of key outcome variables in Study 1.
\begin{table}[]
\resizebox{\textwidth}{!}{%
\begin{tabular}{@{}llcccccc@{}}
\toprule
             &                     & \multicolumn{2}{c}{\textbf{Partisan Animosity}} & \multicolumn{2}{c}{\textbf{Feed-Level Satisfaction}} & \multicolumn{2}{c}{\textbf{Time Spent On Feed}} \\ \midrule
             &                     & \textit{\textbf{M}}    & \textit{\textbf{SD}}   & \textit{\textbf{M}}      & \textit{\textbf{SD}}      & \textit{\textbf{M}}    & \textit{\textbf{SD}}   \\ \midrule
Democrats    & Null                & 72.51                  & 22.45                  & -                        & -                         & -                      & -                      \\
             & Content Warning     & 75.30                  & 18.70                  & 4.40                     & 1.73                      & 425.17                 & 598.43                 \\
             & Removal and Replace & 76.45                  & 21.96                  & 5.20                     & 1.94                      & 344.12                 & 311.62                 \\
             & Downranking           & 76.70                  & 21.62                  & 5.04                     & 2.10                      & 446.19                 & 431.40                 \\
             & Engagement          & 82.01                  & 19.71                  & 4.53                     & 1.91                      & 384.85                 & 373.90                 \\
             & Ideology            & 81.20                  & 21.17                  & 4.51                     & 1.88                      & 353.60                 & 318.68                 \\
             & Chronological       & 76.87                  & 23.25                  & 4.77                     & 2.02                      & 478.48                 & 641.19                 \\
             & \textbf{Total}      & \textbf{77.41}         & \textbf{21.48}         & \textbf{4.75}            & \textbf{1.95}             & \textbf{403.84}        & \textbf{460.69}        \\ \midrule
Republicans  & Null                & 66.72                  & 23.44                  & -                        & -                         & -                      & -                      \\
             & Content Warning     & 74.70                  & 20.89                  & 4.58                     & 1.91                      & 463.20                 & 573.34                 \\
             & Removal and Replace & 70.10                  & 25.03                  & 5.20                     & 1.99                      & 322.66                 & 315.94                 \\
             & Downranking           & 68.03                  & 24.37                  & 4.43                     & 1.84                      & 552.57                 & 1676.04                \\
             & Engagement          & 72.44                  & 18.61                  & 4.52                     & 1.82                      & 429.10                 & 336.28                 \\
             & Ideology            & 71.23                  & 24.38                  & 5.18                     & 2.07                      & 412.51                 & 374.25                 \\
             & Chronological       & 68.19                  & 22.26                  & 4.63                     & 2.03                      & 463.20                 & 573.34                 \\
             & \textbf{Total}      & \textbf{70.20}         & \textbf{22.79}         & \textbf{4.47}            & \textbf{1.81}             & \textbf{432.33}        & \textbf{791.83}        \\ \midrule
Both Parties & Null                & 70.13                  & 22.98                  & -                        & -                         & -                      & -                      \\
             & Content Warning     & 75.04                  & 19.62                  & 4.47                     & 1.81                      & 441.49                 & 586.56                 \\
             & Removal and Replace & 74.37                  & 23.13                  & 5.20                     & 1.95                      & 337.16                 & 312.36                 \\
             & Downranking           & 73.25                  & 23.09                  & 4.80                     & 2.02                      & 488.52                 & 1106.00                \\
             & Engagement          & 78.68                  & 19.82                  & 4.53                     & 1.87                      & 400.24                 & 361.05                 \\
             & Ideology            & 77.46                  & 22.88                  & 4.76                     & 1.97                      & 375.73                 & 340.88                 \\
             & Chronological       & 73.63                  & 23.21                  & 4.71                     & 2.02                      & 445.22                 & 563.02                 \\
             & \textbf{Total}      & \textbf{74.67}         & \textbf{22.26}         & \textbf{4.74}            & \textbf{1.95}             & \textbf{414.51}        & \textbf{606.08}        \\ \bottomrule
\end{tabular}}
\vspace{0.05in}
\caption{Means and $SD$s of Main Outcome Variables}
\label{tab:mean}
\vspace{-0.2in}
\end{table}

\subsection{Prompts Used in Study 2}
\label{appendix:prompts}
Detailed below are the prompts provided to the large language model (GPT-4) for Study 2 and Study 3 for the eight anti-democratic attitude variables.

\subsubsection{V1.} Partisan Animosity

\begin{lstlisting}[language=Markdown]
Please rate the following message's partisan animosity from 1 to 3. Partisan animosity is defined as "dislike for opposing partisans". Your rating should consider whether the following factors exist in the following message:

A: Partisan name-calling
B: Emotion or exaggeration
    
Rate 1 if neither factor exists
Rate 2 if one of the factors exists
Rate 3 if both factors exist
    
After your rating, please provide reasoning in the following format: 
Rating: ### Reason: (### is the separator)
\end{lstlisting}

\subsubsection{V2.} Support for Undemocratic Practices
\begin{lstlisting}[language=Markdown]
Please rate the following message's support for undemocratic practices from 1 to 3. Support for undemocratic practices is defined as "willingness to forgo democratic principles for partisan gain". Undemocratic practices are undemocratic tendencies or actions such as reducing polling stations in areas that support their opponents, attacking the independence of the judiciary, undermining the free press, and challenging the legitimacy of election results, and encouraging political violence. Your rating should consider whether the following factors exist in the following message:

A: Show support for undemocratic practices
B1: Partisan name-calling
B2: Emotion or exaggeration

Rate 1 if doesn't satisfy any of the factors
Rate 2 if doesn't satisfy A, but satisfies B1 or B2
Rate 3 if satisfies A, B1 and B2

After your rating, please provide reasoning in the following format: 
Rating: ###  Reason: (###  is the separator)
\end{lstlisting}

\subsubsection{V3.} Support for Partisan Violence
\begin{lstlisting}[language=Markdown]
Please rate the following message's support for partisan violence from 1 to 3. Support for partisan violence is defined as "willingness to use violent tactics against outpartisans". Examples of partisan violence include sending threatening and intimidating messages to the opponent party, harassing the opponent party on the Internet, using violence in advancing their political goals or winning more races in the next election.
					
Your rating should consider whether the following factors exist in the following message:

A: Show support for partisan violence
B1: Partisan name-calling
B2: Emotion or exaggeration

Rate 1 if doesn't satisfy any of the factors
Rate 2 if doesn't satisfy A, but satisfies B1 or B2
Rate 3 if satisfies A, B1 and B2

After your rating, please provide reasoning in the following format: 
Rating: ###  Reason: (###  is the separator)
\end{lstlisting}

\subsubsection{V4.} Support for Undemocratic Candidates
\begin{lstlisting}[language=Markdown]
Please rate the following message's support for undemocratic candidates from 1 to 3. Support for undemocratic candidates is defined as "willingness to ignore undemocratic practices to elect inparty candidates". Undemocratic candidates are oftentimes those who support the following undemocratic practices such as reducing polling stations in areas that support their opponents, attacking the independence of the judiciary, undermining the free press, and challenging the legitimacy of election results, and encouraging political violence.

Your rating should consider whether the following factors exist in the following message:

A: Show support for undemocratic candidates
B1: Partisan name-calling
B2: Emotion or exaggeration

Rate 1 if doesn't satisfy any of the factors 
Rate 2 if satisfies A, but not B1 or B2
Rate 3 if satisfies A, B1 and B2

After your rating, please provide reasoning in the following format: 
Rating: ###  Reason: (###  is the separator)
\end{lstlisting}

\subsubsection{V5.} Opposition to Bipartisanship
\begin{lstlisting}[language=Markdown]
Please rate the following message's opposition to bipartisanship from 1 to 3. Opposition to bipartisanship is defined as "resistance to cross-partisan collaboration". Your rating should consider whether the following factors exist in the following message:
A: Any name-calling or terms that reduce trust
B: Emotion or exaggeration

Rate 1 if neither factor exists
Rate 2 if one of the factors exists
Rate 3 if both factors exist

After your rating, please provide reasoning in the following format: 
Rating: ###  Reason:(###  is the separator)
\end{lstlisting}

\subsubsection{V6.} Social Distrust
\begin{lstlisting}[language=Markdown]
Please rate the following message's social distrust from 1 to 3. Social distrust is defined as "distrust of people in general". Your rating should consider whether the following factors exist in the following message:
A: Any name-calling or terms that reduce trust
B: Emotion or exaggeration

Rate 1 if neither factor exists
Rate 2 if one of the factors exists
Rate 3 if both factors exist

After your rating, please provide reasoning in the following format: 
Rating:###  Reason: (###  is the separator)
\end{lstlisting}

\subsubsection{V7.} Social Distance
\begin{lstlisting}[language=Markdown]
Please rate the following message's social distance from 1 to 3. Social distance is defined as "resistance to interpersonal contact with outpartisans". Your rating should consider whether the following factors exist in the following message:

A: Any terms that increase distrust, distance, insecurity, hate, prejudice, or discrimination
B1: Emotion or exaggeration
B2: Any events that damages communities or decrease societal trust such as mass shooting 

Rate 1 if doesn't satisfy any of the factors 
Rate 2 if satisfies A, but not B1 or B2
Rate 3 if satisfies A, B1 and B2

After your rating, please provide reasoning in the following format: 
Rating:###  Reason: (###  is the separator)
\end{lstlisting}

\subsubsection{V8.} Biased Evaluation of Politicized Facts
\begin{lstlisting}[language=Markdown]
Please rate the following message's biased evaluation of politicized facts from 1 to 3. Biased evaluation of politicized facts is defined as "skepticism of facts that favor the worldview of the other party". Your rating should consider whether the following factors exist in the following message:

A: partially present political facts or discuss a controversial issue with a certain political stance
B: emotion/exaggeration

Rate 1 if neither factor exists
Rate 2 if one of the factors exists
Rate 3 if both factors exist

After your rating, please provide reasoning in the following format: 
Rating: ###  Reason:(###  is the separator)
\end{lstlisting}

\subsection{Additional Study 2 Results Using GPT-3.5}
\label{appendix:gpt-3.5}
Table~\ref{tab:gpt_3.5_allVars} reports Study 2 results for the GPT-3.5 model from OpenAI, the model variant preceding GPT-4, to explore how LLM-based ranking varies across models. The same prompts, dataset, and study procedures were used to generate these results; the only modification was the use of the \texttt{gpt-3.5-turbo} chat completion model to generate ratings.

\begin{table}[!t]
\centering
\resizebox{\textwidth}{!}{
\begin{tabular}{lccc}
\toprule
\textbf{Individual anti-democratic attitude variables}              & \textbf{Krippendorff's $\alpha$} & \textbf{Classification Accuracy} & \textbf{F1 Score} \\ 
\midrule
Partisan Animosity                     & .694                    & .865                    & .667                        \\
Support for Undemocratic Practices     & -.003                   & .815                    & .497                        \\
Support for Partisan Violence          & .356                    & .840                    & .677                        \\
Support for Undemocratic Candidates    & .039                    & .885                    & .373                       \\
Opposition to Bipartisanship           & .621                    & .850                    & .702                        \\
Social Distrust                        & .628                    & .790                    & .676                        \\
Social Distance                        & .637                    & .790                    & .643                       \\
Biased Evaluation of Politicized Facts & .737                    & .790                    & .690                       \\ 
\bottomrule
\\
\toprule
\rowcolor{purple}
\textbf{Outcome variable}              & \textbf{Krippendorff's $\alpha$} & \textbf{Spearman's $\rho$} & \textbf{Mean Absolute Error (MAE)} \\ 
\midrule
\rowcolor{purple}
\textbf{Overall democratic attitude ranking (8-24 scale)}                     & .761                    & .759     & 1.135            \\
\bottomrule
\end{tabular}
}
\vspace{0.05in}
\caption{Performance metrics for GPT-3.5 ratings on the overall democratic attitude ranking and individual anti-democratic attitude variables. Compared to the GPT-4 results, we observe comparable alignment between GPT-3.5 and manual rating results.
}
\label{tab:gpt_3.5_allVars}
\end{table}

\received{July 2023}
\received[revised]{October 2023}
\received[accepted]{November 2023}

\end{document}